\documentclass[fleqn,10pt]{SelfArx} 

\usepackage[english]{babel} 
\usepackage[final]{pdfpages}

\usepackage[T1]{fontenc}
\usepackage[utf8]{inputenc}
\usepackage{authblk}

\usepackage{graphicx,epsfig}

\usepackage{times}
\usepackage{graphics,dcolumn,bm,fleqn,float}
\usepackage{amssymb,amsmath,multirow,rotate,color}
\usepackage{comment} 
\usepackage{cite} 

\usepackage{graphicx,epsfig}
\usepackage{times}
\usepackage{graphics,dcolumn,bm,fleqn,float}
\usepackage{amssymb,amsmath,multirow,rotate,color}
\usepackage{comment} 

\usepackage{color}

\definecolor{color1}{RGB}{0,0,90} 
\definecolor{color2}{RGB}{0,20,20} 

\usepackage{hyperref} 
\hypersetup{hidelinks,colorlinks,breaklinks=true,urlcolor=color2,citecolor=color1,linkcolor=color1,bookmarksopen=false,pdftitle={Title},pdfauthor={Author}}

\JournalInfo{~} 
\Archive{} 

\PaperTitle{Breaking the spell of nestedness} 

\Authors{\normalsize Cl\`audia Payrat\'o Borr\'as\textsuperscript{1,2}, Laura Hern\'andez\textsuperscript{1}, 
and Yamir Moreno\textsuperscript{2,3,4,5}}

\affiliation{\textsuperscript{2}\textit{Laboratoire de Physique Th\'eorique et Mod\'elisation, UMR CNRS, Universit\'e de Cergy-Pontoise, 2 Avenue Adolphe Chauvin, F-95302, Cergy-Pontoise Cedex, France.}}
\affiliation{\textsuperscript{2}\textit{Institute for Biocomputation and Physics of Complex Systems (BIFI), University of Zaragoza, Spain.}}
\affiliation{\textsuperscript{3}\textit{Department of Theoretical Physics, Faculty of Sciences, University of Zaragoza, Spain.}}
\affiliation{\textsuperscript{4}\textit{Complex Networks and Systems Lagrange Lab, Institute for Scientific Interchange, Turin, Italy.}}
\affiliation{\textsuperscript{5}\textit{Complexity Science Hub Vienna, Austria.}}

\newcommand{\keywordname}{Keywords} 

\Abstract{Mutualistic interactions, which are beneficial for both interacting species, are recurrently present in ecosystems. Observations of natural systems showed that, if we draw mutualistic relationships as binary links between species, the resulting bipartite network of interactions displays a widespread particular ordering called nestedness~\cite{bascompte2003nested}. On the other hand, theoretical works have shown that a nested structure has a positive impact on a number of relevant features ranging from species coexistence~\cite{bastolla2009architecture}, to a higher structural stability of communities and biodiversity~\cite{thebault2010stability, rohr2014structural}. However, how nestedness emerges and what are its determinants, are still open challenges that have led to multiple debates to date~\cite{burgos2007nestedness, suweis2013emergence, valverde2016nestedness}.
Here, we show, by applying a theoretical approach to the analysis of 167 real mutualistic networks, that nestedness is not an irreducible feature, but a consequence of the degree sequences of both guilds of the mutualistic network.
Remarkably, we find that an outstanding majority of the analyzed networks does not show statistical significant nestedness. These findings point to the need of revising previous claims about the role of nestedness and might contribute to expand our understanding of how evolution shapes mutualistic interactions and communities by placing the focus on the local properties rather than on global quantities.}

\begin{document}

\flushbottom 

\maketitle 

\thispagestyle{empty} 

Nested patterns are ubiquitous in ecological systems. This observation has triggered an intense research aimed at defining and measuring nestedness~\cite{ulrich2009consumer} as well as at explaining its origin~\cite{vazquez2006community, rezende2007effects, perazzo2014study}. The interest in deciding whether a system is nested or not goes beyond characterizing it from a merely topological viewpoint. Admittedly, it has been suggested that nestedness plays an important role in biodiversity persistence, a claim which is nevertheless the subject of an ongoing and intense debate ~\cite{bastolla2009architecture, thebault2010stability, rohr2014structural}. Furthermore, the relevance of nestedness as a suitable indicator to characterize mutualistic ecosystems has been recently challenged~\cite{staniczenko2013ghost, james2012disentangling}. Several works have proposed alternative properties of the observed networks to link relevant structural characteristics with the system's dynamics, in particular, the networks' assortativity or the heterogeneity in the number of interactions of the species~\cite{james2012disentangling, jonhson2013factors, feng2014heterogeneity}. 

A key question is then whether nestedness, conceived as a global trait of the emerging architecture, is actually relevant and informative of the ecosystems' dynamics, or contrarily, it just derives from lower order properties of the interaction network. In addition, elucidating the previous question would also allow to solve another open challenge, namely, which is the right \textit{null model} against which one should assess nestedness? The latter is a relevant issue by itself, as any claim concerning statistical significance of a nested pattern implicitly involves its comparison with a null hypothesis (model). In order to address the aforementioned questions, we analyze an empirical set of $167$ mutualistic networks (see \textit{Methods}) to determine if, indeed, the observed amount of nestedness in real ecosystems could solely arise from the empirical degree sequences. Our choice is rooted on a a theoretical work~\cite{medan2007analysis} that showed that the geometric curve that delimits the region with interactions in an ideally nested matrix~\cite{atmar1993measure} can be ultimately related, by means of an approximation, to the \textit{degree distributions} of both guilds of the corresponding bipartite network. 

We constructed a grand canonical ensemble for each empirical ecological web under the constraint that, for the two guilds, the degree sequences in the ensemble match on \textit{average} the empirical ones (see \textit{Methods} and the \textit{SM}). This methodological approach has the advantage that possible missing links or overrated interactions, that have been suggested to lead to impoverished ecological data~\cite{olesen2010missing}, are dealt with in a proper way. In fact, constraining the randomized degree sequences to be equivalent to the empirical ones only on average limits the possible effects of noisy data, while assuring that results are not dependent on specific details. At variance with previous works that have imposed similar constraining rules~\cite{joppa2010nestedness}, here we apply a recently introduced randomizing scheme~\cite{squartini2011analytical, saracco2015randomizing} that treats the ensemble from a statistical physics perspective, yielding the maximum entropy network ensemble such that the degree sequence of the empirical network is found with maximum likehood (see \textit{Methods}). This 

\onecolumn

\noindent enforces, as aimed, the ensemble's mean degree sequences to be the empirical ones whilst precluding common biases of other sampling techniques~\cite{garlaschelli2008maximum}, and allows us to obtain the probability that two potential partners interact in the randomized ensemble, see Fig.1. Moreover, one can explicitly write expressions for the main statistical moments of any network property that can be analytically formulated in terms of the elements of the bipartite adjacency matrix. Since the well-known NODF metric for nestedness fulfills such a condition, we have derived the analytical expressions for the mean and the standard deviation of the nestedness' distribution, see \textit{Methods}.

\begin{figure}
\begin{center}
\includegraphics[width=0.7\textwidth]{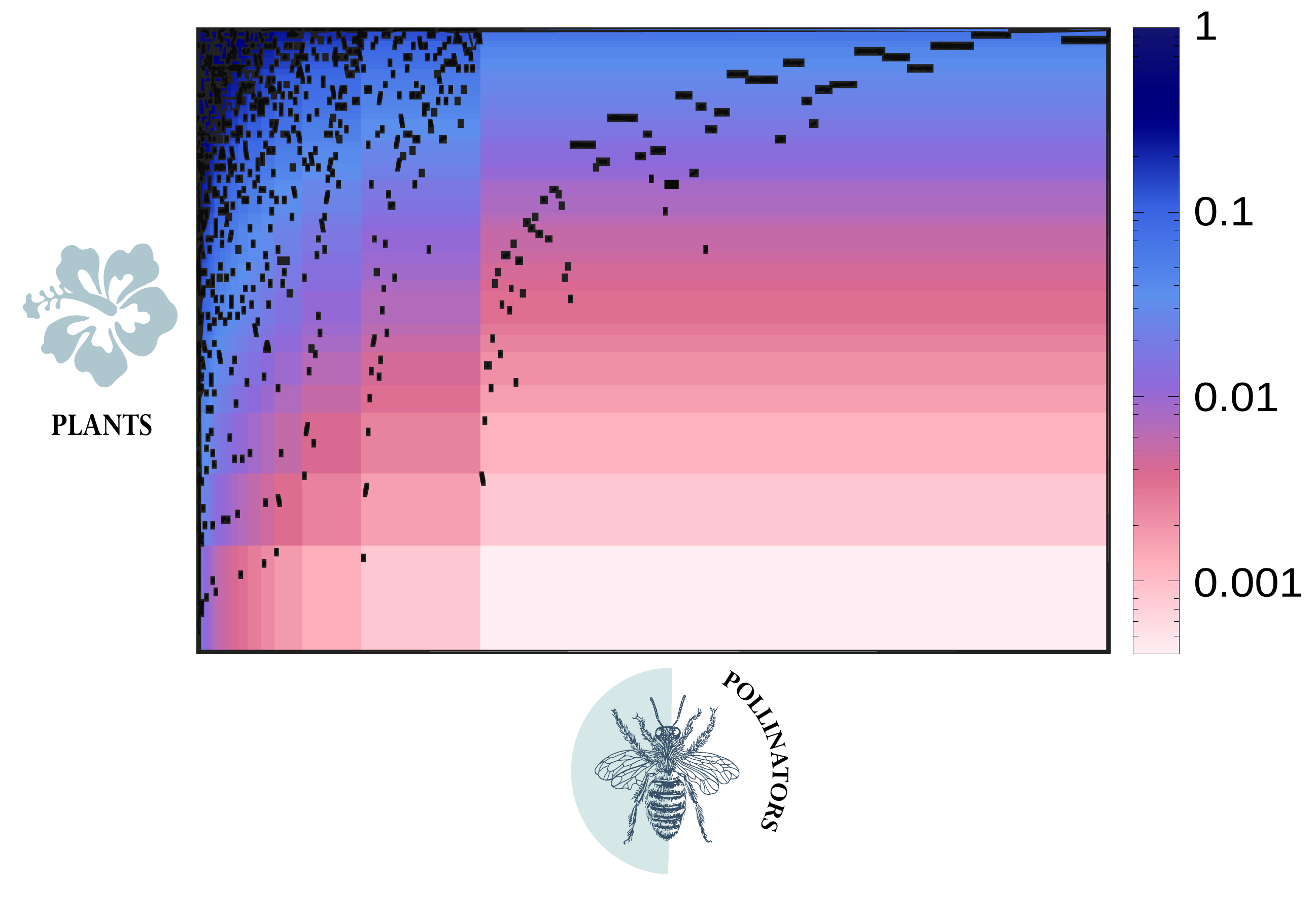} 
\caption{{\bf Comparison of generated and empirical mutualistic interactions.} Probability of interaction between species in the statistical ensemble (color coded as indicated), for the plant-pollinator network recorded by Inoue et al.~\cite{inoue1990insect}. The empirical corresponding bipartite matrix of interactions is superimposed in black. Both plants and pollinators species have been ordered in decreasing order of their degrees (from top to bottom and from left to right).  As it can be seen at a glance, the obtained probabilities are consistent with the observed interactions, with the dark regions delimiting an upper left triangle, as in an ideally nested structure. Note that the color legend is in logarithmic scale.} \label{fig1}
\end{center}
\end{figure}

For each one of the 167 empirical networks we have obtained the interaction probability between the elements of the corresponding bipartite matrix. These networks include three different kinds of mutualistic communities: plant-pollinator, seed-disperser and plant-ant (see Section SI5 of the \textit{SM}). A comparison between the obtained mean expectations of the nestedness of the randomized ensembles and the measured values of the nestedness of the real networks unveils a striking agreement, see Fig. 2. As reported in Table 1, the absolute difference between these two quantities is less than \emph{one} standard deviation for 100 out of 167 networks (59.9\%), raising to 158 out of 167 networks (94.6\%), if we account for \emph{two} standard deviations. The previous percentage increases further after performing a multiple testing correction (see \textit{Methods}): we find that only 3 out of the 167 empirical observations of nestedness are significant ($p$-value $< 0.05$). The three of them, which are of a relatively small size ($\leq 55$ species), were found to be \textit{less} nested than predicted by the statistical ensemble. Additionally, in order to ensure that our findings are not an artifact of using the NODF metric, we have performed the same analysis by using the largest eigenvalue radius, which has been recently proposed as an alternative way of measuring nestedness~\cite{staniczenko2013ghost}. In this case, since it is not possible to obtain an analytical and derivable expression of the metric, we broadly sampled the statistical ensemble using the calculated probabilities and performed the statistical measures on the obtained samples (see \textit{Methods}). This supplementary analysis produces results that are in agreement with those reported above for the NODF, with only 16 out of 167 networks unexpectedly nested (see Section SI3 of the \textit{SM}).

\begin{figure}[!t]
\centering
\includegraphics[width=0.90\textwidth]{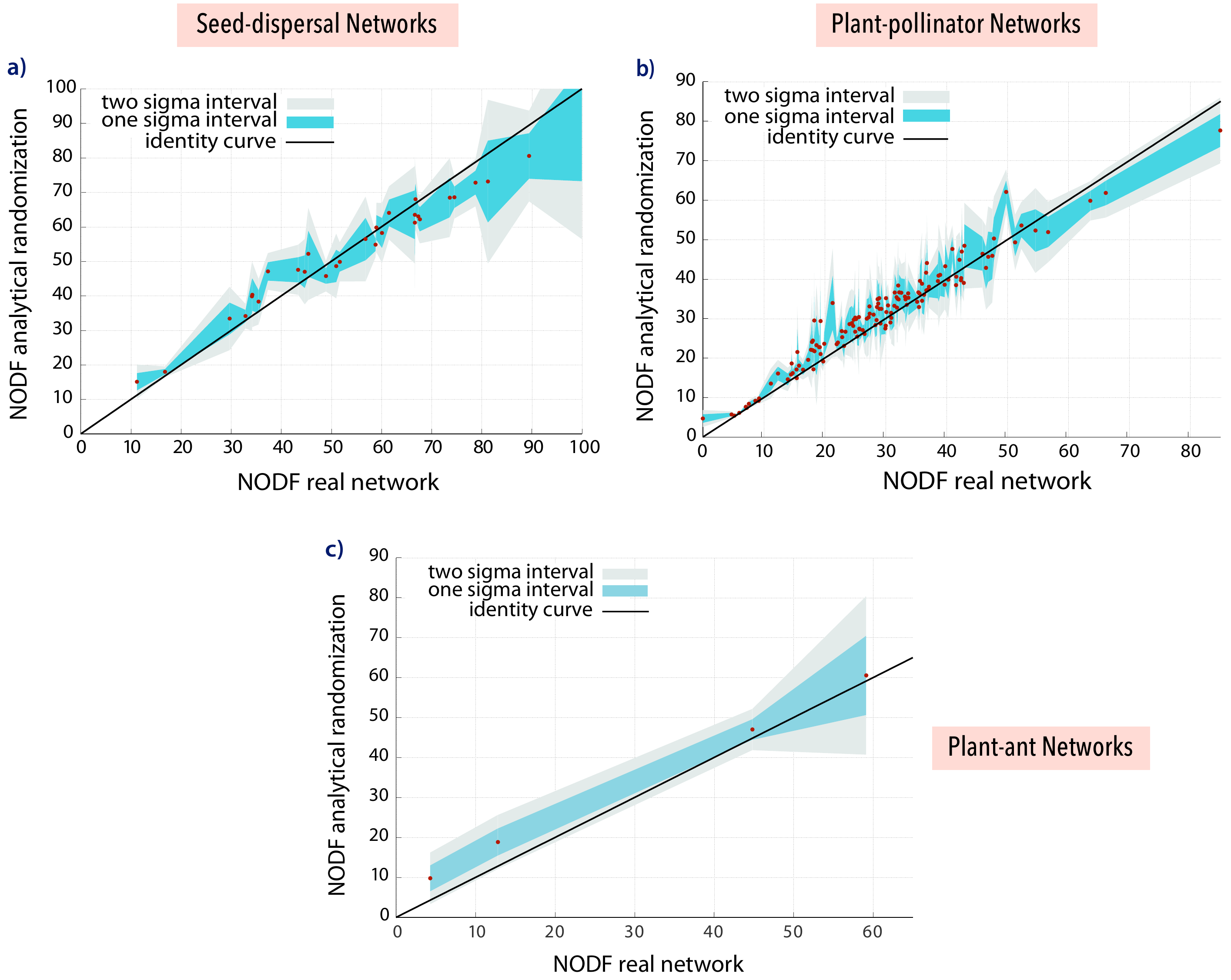} 
\caption{{\bf Significance of the nestedness of real networks.} The figure shows the average value of the nestedness in the generated statistical ensemble against the empirical measure of nestedness (black dots) for the 167 real mutualistic networks. The 3 panels correspond to different kinds of mutualistic systems as indicated. The shadowed areas represent one (teal) and two (light gray) standard deviations of the mean. Further details about the number of networks whose nestedness are within these boundaries are provided in Table 1. A detailed significant test results in only 3 networks having a statistical significant (in all cases under represented) nestedness value. Overall, the results indicate that the nestedness of these mutualistic networks is not significant.}
\label{fig2}
\end{figure}

\begin{center} 
\begin{table}[!b]
\centering
\begin{tabular}{ |c|c|c|c|c|} 
\hline
Type of community &\multicolumn{2}{c|}{fraction of ntws with $\vert$z-score$\vert$ $\leq$ 1} & \multicolumn{2}{c|}{fraction of ntws with $\vert$z-score$\vert$ $\leq$ 2} \\ 
\hline
  \hline 
 plant-pollinator & $\quad$ 82 out of 133 $\quad$ & 61.7\% & $\quad$ 126 out of 133 $\quad$ & 95.5\%  \\ 
 seed-disperser & $\quad$ 16 out of 30 $\quad$ & 53.3\% & $\quad$ 28 out of 30 $\quad$ & 93.3\%  \\  
 plant-ant & $\quad$ 2 out of 4 $\quad$ & 50.0\% & $\quad$ 4 out of 4 $\quad$ & 100.0\% \\
 \hline 
\end{tabular}
\caption{Results, disentangled into communities, showing the fraction of networks (abbreviated above as 'ntws') whose discrepancy between the real and randomized nestedness is less or equal than one or two standard deviations.}
\label{table1}
\end{table}
\end{center} 

The findings above are of utmost importance in at least two fundamental aspects. Firstly, they demonstrate that, given the degree sequence of real networks, the observed nestedness is not significant. Secondly, they show that nestedness is not an irreducible pattern, in sharp contrast to the widely extended belief that it is so. In other words, these results reveal that the observed nested structure of the ecological communities studied is, in fact, a mere consequence of the degree sequences of the two guilds. Moreover, regarding recent debates about the use of a proper null model for nested networks~\cite{ulrich2007null}, our findings point out the need of incorporating the information contained in the degree sequences. Indeed, our results indicate that an appropriate null model is the set of exponential random graphs for which the probability of having the same degree sequences of both branches of the bipartite graphs when compared to a real mutualistic network is maximized. Thus, we propose that the methodology implemented here to obtain the statistical ensemble of graphs that are compatible with the real networks could be a general tool to assess nestedness' significance. 

\begin{figure}
\centering
\includegraphics[width=0.9\columnwidth]{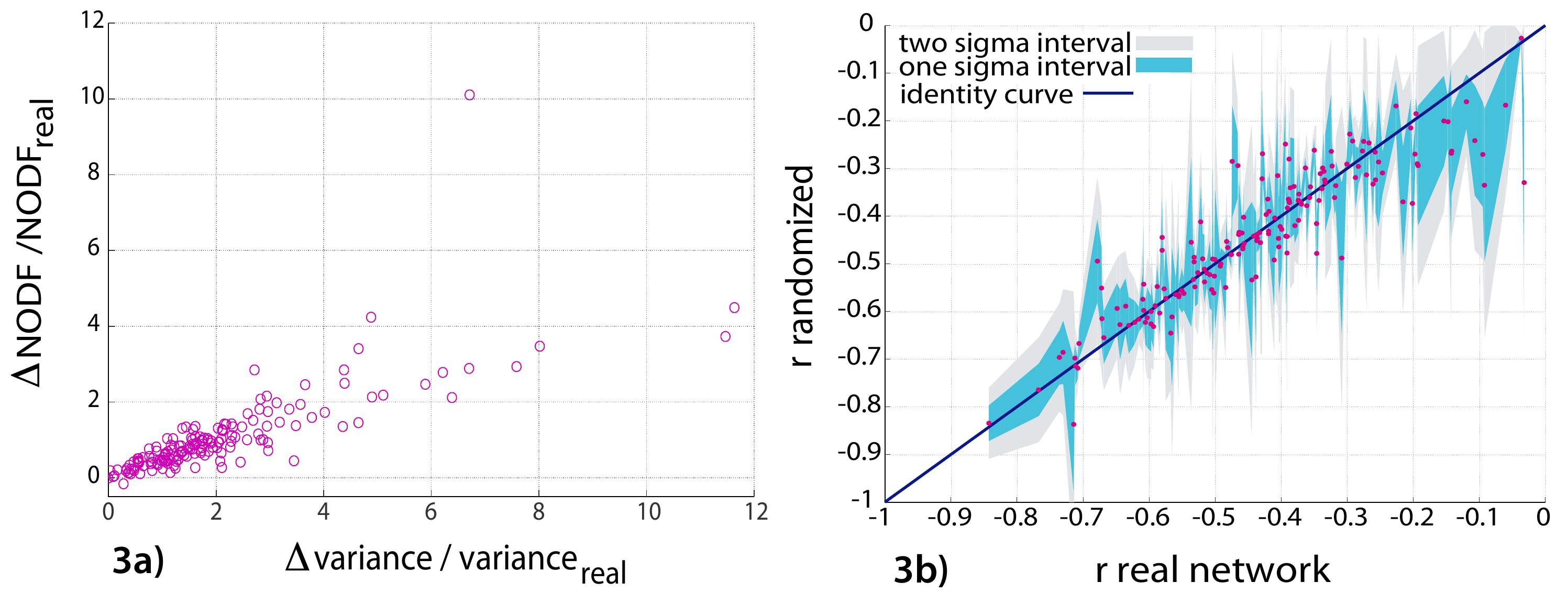}
\caption{{\bf Determinants of nestedness.} Panel \textit{(a)}: relative change in nestedness and the corresponding change in heterogeneity, measured for the set of 167 empirical networks and the average over the respective rewired ones. We used the rewiring algorithm described in \textit{Methods}. Nestedness is measured using the NODF metric, whereas the heterogeneity is measured through the variance of the degree sequence of the unipartite adjacency matrix. We found a correlation index for a linear fit (excluding the top outlier) of $R=0.88$. This closely linear relationship discovers a tight bound between nestedness and heterogeneity. Panel \textit{(b)} shows a comparison between the real observation of the degree assortativity $r$ (Pearson's coefficient among degrees) and the average estimation in the statistical ensemble, for the 167 networks of our study. The fact that $r < 0$ for all values indicate that both real networks and the average of the randomized ensemble are naturally disassortative.} \label{fig3}
\end{figure}

In the light of the previous results, the second question arising is whether we can determine which characteristic of the degree sequences modulates how nested a network is. Considering that the degree distributions of mutualistic communities have been reported to commonly follow a (truncated) power-law~\cite{jordano2003invariant}, we propose, as a plausible candidate, the \textit{heterogeneity} in the number of contacts per species. Thus, our hypothesis is that for two networks with identical number of species and connections but diverse degree sequences, the most heterogeneous one (taking into account both guilds) will be as well the most nested. To evaluate this  conjecture, we made use of a self-organizing network model that is devised with the aim of optimizing the nestedness of a network~\cite{burgos2007nestedness} by rewiring existing links (see \textit{Methods}). After applying this algorithm to our empirical set of networks, we found that the resulting degree sequences are, with respect to the original ones, more heterogeneous and that the final networks are more nested, see Fig. 3a. This allows to bridge the gap between two topological features that have been classically treated separately, though previous works already suggested their connection~\cite{perazzo2014study,jonhson2013factors}. Interestingly enough, the relationship between network's heterogeneity and nestedness also explains why dynamical implications once attributed to nestedness like the sustainability of communities with a large number of different coexisting species~\cite{bastolla2009architecture} or the network's structural stability~\cite{thebault2010stability,memmott2004tolerance}, have recently been successfully associated with other properties such as the heterogeneity itself~\cite{feng2014heterogeneity} or the species' degree~\cite{james2012disentangling}. 

Moreover, accounting for the heterogeneity offers some further insight on the process of emergence of nestedness out of the degree sequences. At first glance, it might not be evident why our null model reproduces so well the empirical nestedness. A priori, we would naively expect that the random ensemble contains both nested and non-nested structures alike, in which specialists appear attached, respectively, to generalists or to other specialists. Although a given number of connections are certainly imposed by the existence of super-generalists as well as by finite size effects, normally there is still room for reshuffling links (like in the "swapping algorithm"~\cite{gotelli2001swap}). In terms of mixing, we would say that, concerning specialists, both assortative configurations (nodes have neighbors with degrees similar to their own) and disassortative ones (neighbors have dissimilar degree) are in theory feasible. Why, then, our algorithm is expected to generate disassortative networks (see Fig. 3b)? Here, the particularity that we used a maximally-entropic ensemble plays a crucial role. Johnson et al.~\cite{johnson2010entropic} showed that, in the case of heterogeneous systems, disassortativity is generally more entropic, that is, it is more likely as long as no external pressures are at work. This occurs, to put it simply, because for a species with few interactions there exist many more chances to engage with another species with numerous connections than matching to a low-connected partner. Therefore, the low significance of empirical nested patterns reported here is directly related to the fact that the number of mutualistic interactions per species is a highly heterogeneous quantity. 
  
In concluding, it is worth mentioning that in recent years, nestedness has been proposed to arise either as an ecological feature that provides an optimal balance between competition and mutualism ~\cite{bastolla2009architecture,suweis2013emergence}, or as a byproduct of processes such as the assembling rules~\cite{takemoto2010nested,valverde2016nestedness}. Our results imply that no selective pressure has acted upon nestedness, which does not exclude, however, that such pressure has shaped the degree sequences. Even though such conclusions do not invalidate nestedness' usefulness as an indicator of stability or robustness, we would like to underline that our findings clearly demonstrate that the degree sequences are the lower-order determinants of nestedness. Moreover, this highlights the interest of focusing on the ecological and evolutionary mechanisms that have led to the heterogeneous degree distributions present in mutualistic ecosystems~\cite{johnson2000generalization,bronstein2006evolution}, like might be the need to diminish the cost of mutualism~\cite{bronstein2001costs}. Understanding the way in which structural properties emerge in ecological communities is a fundamental, long-standing challenge that can provide critical clues to depict ecosystems' past assembling, present functioning and future responses. Finally, given that nested patterns have been recurrently detected across systems as diverse as biological, social and technological networks, our findings are expected to have relevant implications beyond the present analysis of ecological mutualistic communities. 

\section*{Methods}

\paragraph{Construction of the Random Ensemble.} We constructed an ensemble following the Exponential Random Graph model. This ensemble maximizes the Shannon-Gibbs Entropy given the average degree sequences of the two guilds of a bipartite network as constraint. Yet, it is not fully determined due to the presence of some free Lagrange multipliers resulting from the constrained optimization. Following Squartini and Garlaschelli~\cite{squartini2011analytical}~\cite{garlaschelli2008maximum}, we imposed that the degree sequences of the empirical network are found with maximum likelihood. This provides a set of coupled equations to solve for the Lagrange multipliers (one equation per node, see Eqs.~7-8 in Section SI1 of the \textit{SM}). Determining the statistical random ensembles of our 167 empirical networks entails solving computationally 167 optimization problems. For each network, we numerically found the Lagrange multipliers that maximize the likelihood using two different, independent algorithms: \textit{1)}~a global, pseudo-random numerical method for optimizing the likelihood and \textit{2)}~a deterministic, gradient-based algorithm for solving non-linear systems of equations. See Section SI1 of the \textit{SM} for the explicit expressions of the ensemble probability and equations to solve, as well as more information on the numerical implementation. As shown in~\cite{squartini2011analytical}, in the case of local constraints (as the degree sequences), the probability of existence of a graph in the ensemble can be exactly factorized into the probabilities of existence of a link between species~\cite{squartini2011analytical}. Therefore, after numerically determining each optimal set of Lagrange multipliers, we built the matrix containing the average probability of interaction corresponding to each empirical network (see an example in Fig.~1 and Eqs.~9-10 in Section SI1 of the \textit{SM}).  

\paragraph{Statistical Measures on the Random Ensemble.} We performed the statistical measures on the ensemble following either of the two following approaches. On the one hand, as long as the property that we aim to evaluate could be formulated as an analytical and derivable expression, Squartini and Garlaschelli showed~\cite{squartini2011analytical} that it is possible to obtain, at first order, the analytical expression of the first and second moments of the corresponding distribution. These expressions depend only on the link probabilities (see Eq. 1-2 in Section SI2 of the \textit{SM}). We wrote the NODF metric in a compact, analytical form and derived the expression of the theoretical average expectation and standard deviation of nestedness in the ensemble (Eqs.~7-9 in Section SI2 of the \textit{SM}). Finally, we obtained the 167 probability matrices of interactions and computed the main statistical moments. On the other hand, one can always sample the ensemble in order to study the statistics of the target property on a generated sampling. Using this scheme, we produced $10^4$ networks that were assembled using the obtained probability matrix of interactions. Over this subset, we numerically calculated the average expectation and the standard deviation of the largest eigenvalue radius~\cite{staniczenko2013ghost} (see Section SI3 of the \textit{SM}) and the assortativity index measured through the Pearson coefficient of the degrees (see Section SI4 of the \textit{SM}). 

\paragraph{Significance tests.} We quantified the significance of the nestedness using the \textit{z-score} index, which for a general property $x$ reads: $\frac{x^{*}- \langle x \rangle}{\sigma_x}$. For us, $\langle x \rangle$ is the average nestedness computed in the ensemble, and we compare it with the empirical observations $x^{*}$. The standard deviation is $\sigma_x$. Given that the NODF values are gaussian distributed in the random ensemble (see Section SI2 of the \textit{SM}), the $z$-scores can be directly related to $p$-values. We performed a \textit{multiple test correction} which allows accounting for the fact that as the number of statistical tests increases, so does the probability of finding rare events~\cite{benjamini1995controlling}. Thus, when considering the multiple comparisons we could prevent overstating the number of significant discoveries. It is pertinent to apply this technique here since the 167 cases studied are evaluated under the same \textit{null hypothesis} and all of them follow a normal distribution. We employed the \textit{false discovery rate} method, in particular the Benjamini-Hochberg procedure which applies to independent tests~\cite{benjamini1995controlling}.

\paragraph{Self organizing network model.} In order to reorganize the original network into an even more nested structure, we numerically implemented the self-organizing network model proposed by Burgos et al.~\cite{burgos2007nestedness}. This methodology keeps constant many aspects susceptible to affect the measure of nestedness, like the size and fill, but modifies the degree sequences through the redistribution of connections. We rewired the links among species following two simple rules: \textit{i)} when changing an interaction, the new partner must have higher degree than the old neighbor \textit{ii)} if the proposed redistribution leaves one of the two nodes with no interactions at all, we reject the change. This operation was repeated until the system achieved a frozen state in which no more reconnections were accepted (we considered this happened when $10^3 N$ consecutive rejections occurred, being $N$ the number of nodes of the network). The final frozen state is normally not perfectly nested, since condition \textit{ii} typically leads to configurations which are not utterly optimal. To compensate this, we carried out $10^3$ independent rewiring operations for each network. We then averaged the target properties, namely, nestedness (measured using NODF) and the variance of the joint degree sequence of the two guilds. 




\section*{SUPPLEMENTARY INFORMATION}

\subsection*{S1. Construction of the random ensemble}

This Section provides additional details on how we constructed the statistical ensembles. An ensemble is a set of networks across which unconstrained features will vary randomly, and over which we will perform statistical measures. 

\subsubsection*{General randomizing scheme}

We denote a network in the ensemble by its graph \textbf{G}, except for the real network which is \textbf{G$^{*}$}. We characterize the ensemble by the probability of occurrence of each of its elements, $P(\textbf{G})$. Following~\cite{park2004statistical} and~\cite{squartini2011analytical}, we choose a probability such that the constraints are only satisfied on average, thus allowing slight mismatches across the ensemble. This is equivalent to constructing a \emph{grand-canonical ensemble} (as opposite to the \emph{micro-canonical ensemble}, where the constraints need to be exactly met always). 

As proposed in \cite{park2004statistical}, we ask the probability of each graph in the ensemble to maximize the Shannon-Gibbs entropy, defined as: 
\begin{equation}
S= -\sum_{\textbf{G}}{P(\textbf{G}) ln P(\textbf{G})}, \label{eq0}
\end{equation}
where the sum runs over all the graphs $G$ in the ensemble. This leads to the \emph{Exponential Random Graph model}, which reads:
\begin{equation}
P(G/\vec{\theta}) = \frac{e^{-H(G,\vec{\theta})}}{Z(\vec{\theta})},\label{eq01}
\end{equation}

being $H$ the graph Hamiltonian such that $H(G, \vec{\theta}) = \vec{\theta} \cdot \vec{C(G)}$, and $Z$ the normalizing partition function $Z=\sum{ e^{-H(G)}}$. The set of variables $\vec{\theta}$ are the \emph{Lagrange multipliers}, resulting from the maximization of Eq.~\ref{eq0} under the chosen constraints $\vec{C}$.

Secondly, we proceed to calculate the exact values of the Lagrange multipliers. Following Squartini and Garlaschelli~\cite{squartini2011analytical}~\cite{garlaschelli2008maximum}, we determine these parameters by imposing that the real network is found in the ensemble with maximum probability. Indeed, we may write the log-likelihood of observing the real network $L(\vec{\theta})  = ln ( P(G^{*} \mid \vec{\theta}))$ as:  

\begin{equation}
L(\vec{\theta}) = -H(G^{*}, \vec{\theta}) - ln Z(\vec{\theta}).   
\end{equation}

Maximizing this quantity thus allows fixing the $\vec{\theta^{*}}$ values. This second requirement ensures not only that the constraints are met on average, but also that they are the most likely ones, which is a warranty of non-bias~\cite{squartini2011analytical}.

\subsubsection*{Ensemble for a bipartite network with constrained degree sequences}

We now explain how the randomizing scheme by Squartini and Garlaschelli~\cite{squartini2011analytical} applies to our specific problem, namely a \emph{bipartite network} subject to \emph{local} constraints. The scheme has already been applied to study international trade networks \cite{saracco2015randomizing}.

To begin with, we construct the hamiltonian for a bipartite network, whose bipartite matrix we call $\textbf{B}$. At variance with the monopartite case, we have \emph{two} degree sequences (one for each of the guilds) which need to be taken into account separately. Although the scheme is equally valid for any mutualistic network (seed-dispersers, ant-plants...), for the sake of clarity we restrict our notation to the paradigmatic case of plant-pollinator communities. Thus, we will speak of systems with $N_P$ number of \emph{plants} and $N_A$ pollinating \emph{animals}. The constrained degree sequences, given by the real network, will be represented respectively by $\vec{v}$ and $\vec{h}$, where $v_p$ is the diversity of \emph{visiting} animal species that a plant species $p$ receives, while $h_a$ is the number of different $hosting$ plant species with which a pollinator species $a$ interacts. 

In order to enforce both distributions as constraints we introduce two sets of Lagrange multipliers,  $\vec{\alpha}$ for plants and $\vec{\beta}$ for animals. Subsequently, the graph hamiltonian can be written as
 
\begin{equation}
H(\textbf{B}, \vec{\alpha}, \vec{\beta}) = \vec{\alpha} \cdot \vec{v} + \vec{\beta} \cdot \vec{h}
\end{equation}

This means that the probability, Eq.~\ref{eq01}, of encountering a bipartite graph $B$ in the exponential random graph ensemble becomes:

\begin{equation}
P(\textbf{B} \mid \vec{\alpha}, \vec{\beta}) = \frac{e^{-\vec{\alpha} \cdot \vec{v} - \vec{\beta} \cdot \vec{h}}}{\sum_{B} e^{-\vec{\alpha} \cdot \vec{v} - \vec{\beta} \cdot \vec{h}} } 
\end{equation}

To simplify the notation we introduce the variable change $x_p = e^{-{\alpha_p}}$ and $y_a = e^{-{\beta_a}}$, as suggested as well by Squartini and Garlaschelli. Then, the log-likelihood of encountering the real network is:

\begin{equation}
L(\vec{x}) = \sum_{p=1} ^ {N_P} v_p \, ln ({x_p}) + \sum_{a=1} ^ {N_A} h_a \, ln (y_a) - \sum_{a=1}^{N_A} \sum_{p=1} ^ {N_P} ln (1\,+\,x_p y_a) \label{eq1}
\end{equation}

which we need to maximize in order to find the optimal variables $\vec{x^*}$ and $\vec{y^*}$, that ultimately define our ensemble. Indeed, by requiring that $\vec{\nabla L} (\vec{x}, \vec{y}) = \vec{0} $, we obtain the following set of equations: 

\begin{align} 
v_p = \sum_{a=1}^{N_A} \frac{x_p y_a}{1+x_p y_a} \; \; \text{ for } p=1, ..., N_P \label{eq2a}\\
h_a = \sum_{p=1}^{N_P} \frac{x_p y_a}{1+x_p y_a} \; \; \text{ for } a=1, ..., N_A \label{eq2b}
\end{align}

It can be easily shown that these equations are equivalent to imposing that the average degrees (right hand side) are equal to the degree sequence from the real network (left hand side), as we do below. 

\subsubsection*{Probability matrix of interactions}

Garlaschelli and Squartini also showed~\cite{squartini2011analytical} that in the case of local constraints, the ensemble probability can be factorized, using our notation, in terms of the \emph{probability of existence of a link between a plant species '$p$' and animal species '$a$'}, which we call $p_{pa}$. In effect, by taking $p_{pa} = \frac{x_p y_a}{1+x_p y_a}$, replacing it into equation 5 and doing some little algebra, one finds: 

\begin{equation}
P(\textbf{B} \mid \vec{\alpha}, \vec{\beta}) = \prod_{p,a} p_{p a} \, ^{b_{pa}} \: (1-p_{pa})^{1-b_{pa}}\label{eq3}
\end{equation}

Where $b_{pa}$ is the ($p$,$a$) element of the bipartite matrix of interactions. Then, using expression~\ref{eq3}, it is almost immediate to see that $ \left\langle b_{pa} \right\rangle=p_{pa}$, thus in turn, $ \left\langle b_{pa} \right\rangle= \frac{x_p y_a}{1+x_p y_a}$. This shows that, as we had said, the right hand-side of equations \ref{eq2a}-\ref{eq2b} is a sum over a column or row of expected values of the randomized bipartite matrix. 

We also note that the possibility of factorizing $P(\textbf{B} \mid \vec{\alpha}, \vec{\beta})$ essentially entails that the probabilities $p_{pa}$ are independent among them. In other words, when the constraints enforced are local, the probability of existence of different links are independent among them. This automatically allows the construction of the exact expected randomized matrix of interactions:

\begin{equation}
  \ \left\langle \textbf{B$^{*}$} \right\rangle =
  \begin{pmatrix}
    p_{1 1} & p_{1 2} & ...& p_{1 a} & ... & p_{1 \textsc{\tiny $N_A$}} \\
   p_{2 1} & p_{2 2} & ...& p_{1 a} & ...& p_{2 \textsc{\tiny $N_A$}} \\
   ... & ... & ... & ... & ... &  ...  \\
   p_{p 1} & p_{p 2} & ...& p_{p a} & ... &  p_{p \textsc{\tiny $N_A$}} \\
   ... & ... & ... &  ... &  ...  &  ...  \\
   p_{\textsc{\tiny $N_P$} 1} & p_{\textsc{\tiny $N_P$} 2} & ... & p_{{\textsc{\tiny $N_P$} a}}& ... & p_{\textsc{\tiny $N_P$} \textsc{\tiny $N_A$}}
  \end{pmatrix}
  \label{eq:matrix1}
\end{equation}

\subsubsection*{Computational implementation}

Here we give the numerical details on how we obtained the Lagrange multipliers $\vec{x}^{*}$ and $\vec{y}^{*}$ that eventually define the corresponding statistical ensembles of the empirical networks. As proposed by Squartini and Garlaschelli \cite{squartini2011analytical}, encountering these multipliers might be achieved following either of two procedures: by directly maximizing the log-likelihood in Eq.~\ref{eq1} through an optimizing algorithm, or by solving the non-linear, coupled set of equations in \ref{eq2a}-\ref{eq2b}. 

First, we optimized the log-likelihood by means of a global search, pseudo-random algorithm belonging to the Monte-Carlo family and known as \textit{simulated annealing} \cite{corana1987minimizing,goffe1994global,goffe1996simann}. In short, this method aims to find the global minimum of a function by sequentially exploring the solution space, through producing random proposals subjected to an accepting criteria. More specifically, it uses the Metropolis criteria \cite{metropolis1953equation}, which is driven by a parameter $T$ traditionally called \textit{temperature}. The algorithm works iteratively, by finding the most probable state at each temperature and then using it as the initial condition in the next step. We start with a high temperature, which means that almost all proposals are accepted. As the algorithm progresses, the temperature decreases and so does the acceptance probability. This makes the search to become more and more restricted around eventual solutions, until a certain number of consecutive iterations (in our case, five) have produced solutions differing less than a certain tolerance, called $tol$. When that happens, we are confident enough of having reached the ground state and the algorithm stops. Given the pseudo-aleatory character of this approach, which allows escaping from local hills, it is extendedly used in situations in which the co-existence of several local optima is suspected.

In our case, since the algorithm is originally intended for minimizing but we aim to maximize, we simply took the minus of the function. Additionally, instead of optimizing the log-likelihood as written in Eq.~\ref{eq1}, we incorporated the fact that the degrees may be degenerate. This means that nodes of the same guild having identical degrees satisfy equivalent equations, hence necessarily bearing the same solution. To account for this, we introduced a multiplicity factor $m_p$ for plants and $m_a$ for animals. If we call $red_P$ and $red_A$ the redundancy for plants and for animals (namely, the corresponding numbers of repeated degrees), then the system can be redimensionalized to $N_P ' = N_P - red_P$ and $N_A ' = N_A - red_A$. Consequently the log-likelihood might be rewritten into: 

\begin{equation}
L(\vec{x}) = \sum_{p=1} ^ {N_P'} m_p v_p \, ln ({x_p}) + \sum_{a=1} ^ {N_A'} m_a h_a \, ln (y_a) - \sum_{a=1}^{N_A'} \sum_{p=1} ^ {N_P'} m_p m_a  \; ln (1\,+\,x_p y_a) \label{eq8}
\end{equation}    

Although, in analytical terms, the original expression in Eq.~\ref{eq1} and this latter one are obviously equivalent, from a computational point of view reducing the number of variables enhances the algorithm's efficiency. Besides, imposing from the beginning such identity between variables improves the accuracy of the program. 

We programmed a classical version of simulated annealing, with a starting temperature of $T=10^3$, a reduction factor of the temperature of $RT=0.85$, a tolerance $tol=10^{-6}$ and a total number of updates per fixed temperature of $2\cdot 10^4$. Furthermore, we ran the algorithm $10$ times per network with different random seeds, in order to produce independent sequences of explorations. When all runs convergence to the same solution, as outlined by Goffe \cite{goffe1996simann}, it is extremely probable that we have certainly encountered the global optimum.  

Secondly, we solved the set of equations by means of a local, deterministic algorithm known as the \textit{modified Powell hybrid method}. In particular, we used the MINPACK library \cite{more1980user} for FORTRAN, available online \cite{minpack}. This method finds the zero of a non-linear system by exploiting its Jacobian, which we analytically calculated and implemented into the program.

Like before, we re-dimensionalized the problem to $N_P'$ equations for plants and $N_A'$ equations for animals, which now read: 

\begin{align} 
v_p = \sum_{a=1}^{N_A'} \frac{ m_a x_p y_a}{1+x_p y_a} \; \; \text{ for } p=1, ..., N_P' \label{eq9a}\\
h_a = \sum_{p=1}^{N_P'} \frac{m_p x_p y_a}{1+x_p y_a} \; \; \text{ for } a=1, ..., N_A' \label{eq9b}
\end{align} 

We implemented these equations and their Jacobian and ran the algorithm with a tolerance $tol=10^{-11}$ (as defined in the source code). The possibility of exploiting the gradient provides, in general, a greater local accuracy than the simulated annealing technique. However, its shortcoming lays in the risk of getting trapped in local optima, from which, due to its deterministic nature, it is unable to escape. To compensate this drawback we performed a significant sampling of the space of initial conditions, by running $10^4$ iterations of the algorithm, each with a different random selection of starting points, covering as well distinct ranges. However, due to the encounter of rough, rather accidental configuration surfaces, the modified Powell hybrid method was not always able to converge to a solution. The rate of success was approximately $50\%$·
 
To finally ensure that we found the global maximum, we compared the outcomes of the various independent runs and also, when the Powell algorithm succeeded, among both methods (so in total $10$ runs for the simulated annealing and $10^4$ for the Powell hybrid method). We confirmed, in all cases, that the same maxima was found. This enables us to assume that we found the actual \emph{global} optimizing Lagrange Multipliers for each one of the networks of our study. 

Moreover, the constraints were successfully met with a relative precision between $0.01\%$ and $10\%$. This check was carried on by computing the average degrees using equations~\ref{eq2a}-\ref{eq2b} and comparing the output with the imposed degree sequences (extracted from the empirical networks). The worst case of $10\%$ was typically caused by discrepancies in low degrees, generally the most sensitive to imprecisions in the elements of the randomized matrix (since the matrix elements of low degree nodes are usually very small, see Fig.1 in main text as an example). Altogether, this second check warrants that our constrained optimization worked as aimed. 
 
\subsection*{S2. Nestedness statistical measures using \textit{NODF}}

Here we describe how to obtain statistical measures in the random ensemble through analytical expressions, and particularly present our derivation for the nestedness metric known as NODF~\cite{almeida2008consistent}.

\subsubsection*{General analytical expressions}
Let us call a property by $X$ and its randomized measure (that is, the average across the random ensemble) by ${\left\langle  {X} \right\rangle} ^{*}$. When the property $X$ can be calculated through an \textit{analytical} expression (non-algorithmic) as a function of the bipartite matrix \textbf{B}, then Squartini and Garlaschelli \cite{squartini2011analytical} showed that it is possible to perform an approximate but accurate measure of the first and second moments of $X$, \emph{directly} on $\left\langle \textbf{B$^{*}$} \right\rangle$ (see Eq.~10, SI1). In particular, for the bipartite case, this reads: 

\begin{equation}
{\left\langle  {X} \right\rangle} ^{*} \simeq X(\left\langle \textbf{B$^{*}$} \right\rangle) \label{eq4}
\end{equation}

\begin{equation}
\sigma_{X} \simeq \sqrt{\sum_{p=1}^{N_P} \sum_{a=1}^{N_A} \left( \frac{\partial X(\textbf{B})}{\partial b_{pa}} \right)^2 \sigma^2 _{b_{pa}}} \label{eq5}
\end{equation} 

Where $\sigma_{b_{pa}}$ is the standard deviation for the bipartite matrix element $b_{pa}$. The condition for these approximations to be accurate is that the property $X$ be gaussian-distributed in the random ensemble.

\subsubsection*{Derivation for \textit{NODF}}

Let us see how the previous expressions can be applied to measure nestedness with the \textit{nestedness metric based on overlap and decreasing fill} (hereafter, \textit{NODF}) by Almeida-Neto et al. \cite{almeida2008consistent}. 

We first verified that the assumption of gaussianity is fulfilled by performing a check on a smaller subset of our set of empirical networks. To do this, for each of the corresponding statistical ensembles we generated a sample of $10^4$ networks obeying the probability of link existence given by $\left\langle \textbf{B$^{*}$} \right\rangle$ (see Eq.~10, SI1). We then computed the nestedness of each sampled network in order to generate the nestedness distribution. In all cases we could successfully fit a gaussian function (see Fig.~\ref{fig2} as an example). 

\begin{figure}[!t]
\centering
  \includegraphics[width=\linewidth]{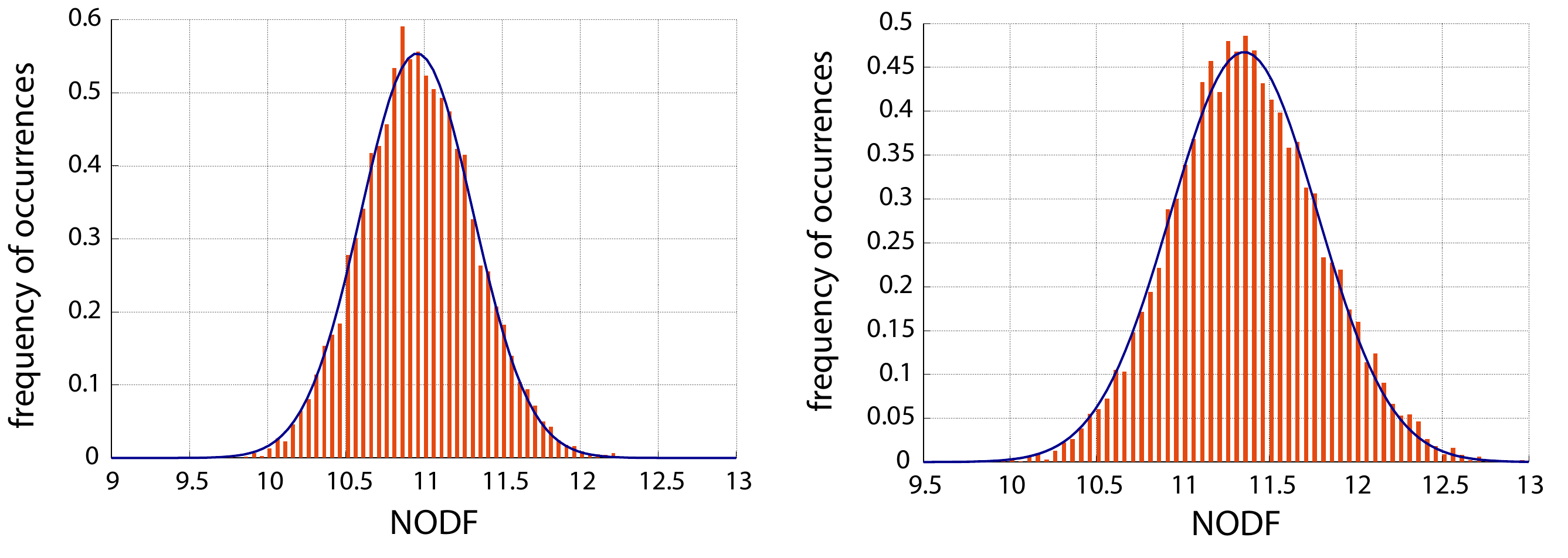}
\caption{Nestedness distribution for two samplings of the statistical ensembles corresponding to the empirical networks by Petanidou et al.~\cite{petanidou1991pollination} (left) and by Inoue et al.~\cite{inoue1990insect} (right). In blue, fits of a gaussian function using the mean and standard deviation extracted from each distribution.}\label{fig2}
\end{figure}

Once shown that the gaussian behavior is satisfied for this metric, we next apply Eqs.~\ref{eq4}-\ref{eq5}. Yet, we need an analytical, packed expression that facilitates the calculations of the metric. The NODF basically considers two contributing factors to nestedness: \emph{decreasing fill} (the fact that, after a proper ordering, both degree sequences progressively decline) and \emph{paired overlap} (the number of shared partners between two columns or rows, normalized by the smaller degree). By gathering together the sequential analysis indicated by Almeida-Neto et al., we proposed a novel compact expression to calculate NODF, that reads:

\begin{equation}
\text{NODF(\textbf{B})} = \frac{1}{K} \; \sum _{i<j}^{N_P} \; \left\lbrace  \left[ 1- \theta(v_{j} - v_{i})\right] \cdot \frac{ \sum\limits_{a=1} ^{N_A} b_{ia} b_{ja} }{v_j} \right\rbrace + \dfrac{1}{K} \; \sum _{k<l}^{N_A} \; \left\lbrace \left[ 1- \theta(h_{l} - h_{k})\right] \cdot \frac{ \sum\limits_{p=1} ^{N_P} b_{pk} b_{pl} }{h_l} \right\rbrace \label{eq6a}
\end{equation} 
 
\begin{align}
\text{where } K^{-1}= \frac{N_P(N_P-1) + N_A(N_A+1)}{200}
\end{align}

We maintain here our previous notation (see SI1), so $v_p$ is the degree of plant $p$ and $h_a$ the degree of animal $a$. The double sums run over two indices such that, as seen in Fig.~\ref{fig2}, row $i$ is placed upper row $j$ and column $k$ more to the left than column $l$. The $K$ factor contains the normalization over all possible pairs, as well as a percentage rescaling (note that NODF takes values between 0 and 100). Finally, the $\theta$ term is the Heaviside step function, which is zero when its argument is negative, and one if its argument is positive or zero. In our context it serves to encapsulate the decreasing fill condition. In fact, from now on we will use the following abbreviations:
\begin{equation}
DF_{i j} =  1- \theta(v_{j} - v_{i}) \quad \text{ such that, if } v_{j} \geq v_{i} \;\text{ then }DF_{i j} =  0,\; \: \text{ and  if } v_{j} < v_{i} \; \text{ then } DF_{i j} =  1 
\end{equation}
\begin{equation}
DF_{k l} =  1- \theta(h_{l} - h_{k}) \quad \text{ such that, if } h_{l} \geq h_{k} \; \text{ then } DF_{k l} =  0, \; \:\text{ and  if } h_{l} < h_{k} \; \text{ then } DF_{k l} =  1 
\end{equation}

\begin{figure}[!tb]
\centering
\includegraphics[width=0.5\columnwidth]{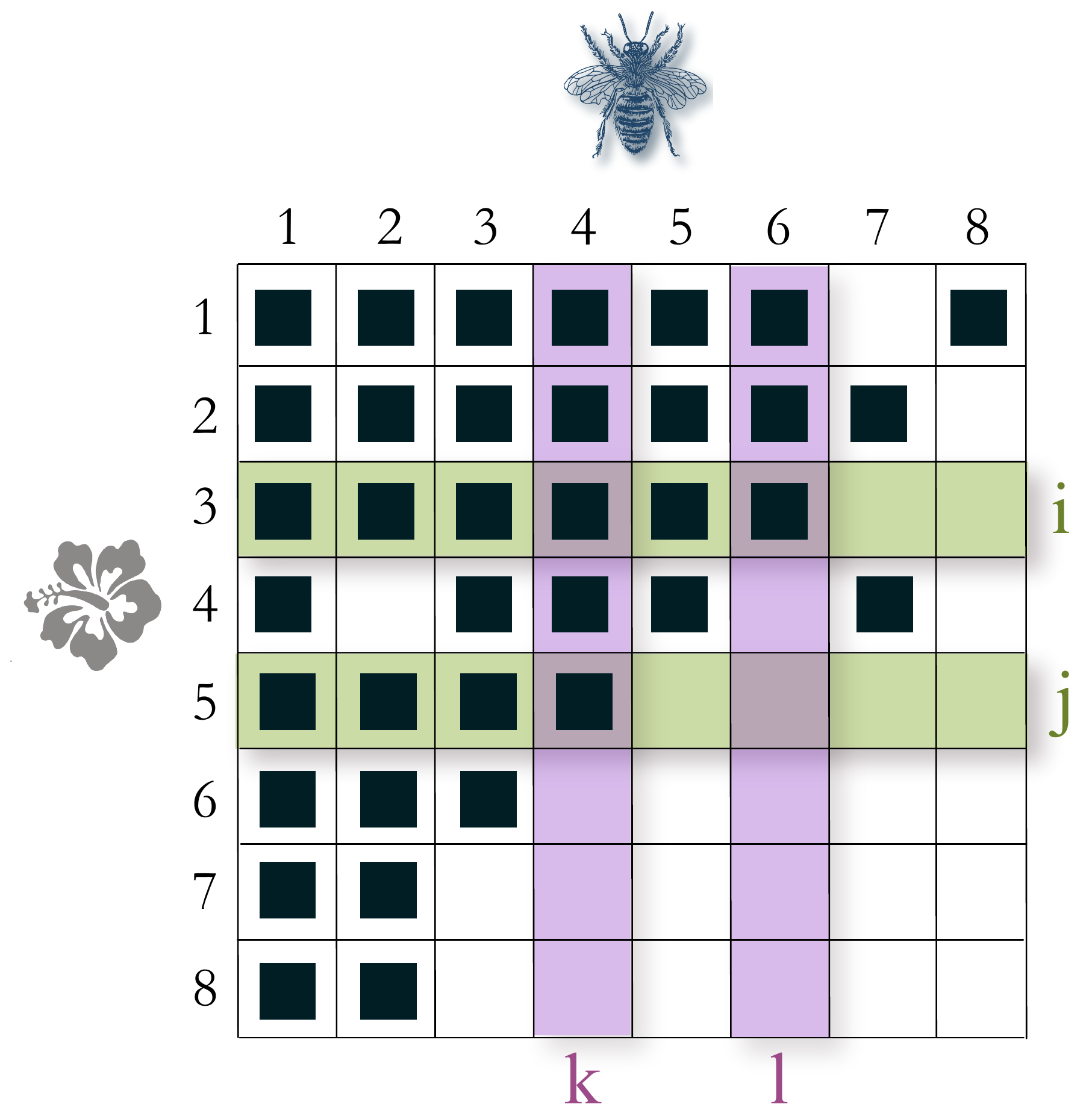}
\caption{Example of an ordered matrix of interactions, not perfectly nested. Species of both guilds have been ordered in decreasing degree, and the numbered labels indicate their rank (the larger the degree, the smaller the rank). The indexes $i$, $j$, $k$ and $l$ illustrate our notation for rows and columns.}
\label{fig2}
\end{figure}

The analytical and packed expression for NODF that appears in Eq.~\ref{eq6a} can then be plugged into Eqs. \ref{eq4}-\ref{eq5}. Accordingly, we obtained that the first moment of the randomized NODF for a certain real bipartite matrix \textbf{B$^*$} reads:

\begin{equation}
\left\langle  \text{NODF(\textbf{B})} \right\rangle ^{*} = \frac{1}{K} \; \sum _{i<j}^{N_P} \; \left\lbrace  DF_{ij} \cdot \frac{ \sum\limits _{a=1} ^{N_A} \left\langle b_{ia} \right\rangle \left\langle b_{ja}\right\rangle  }{\sum\limits_{a=1}^{N_A} \left\langle b_{ja} \right\rangle} \right\rbrace + \frac{1}{K} \; \sum _{k<l}^{N_A} \; \left\lbrace DF_{kl} \cdot \frac{ \sum\limits _{p=1} ^{N_P} \left\langle b_{pk} \right\rangle \left\langle b_{pl}\right\rangle  }{\sum\limits_{p=1}^{N_P} \left\langle b_{pl} \right\rangle} \right\rbrace \label{eq6}
\end{equation} 

Notice that $\sum_{a=1}^{N_A} \left\langle b_{pa} \right\rangle=v_{p}$ and $\sum_{p=1}^{N_P} \left\langle b_{pa} \right\rangle=h_{a}$, given that the randomized matrix necessarily fulfills the enforced constrains. Additionally, this warrants that the ordering of the matrix is equal to the original one, which is important since NODF is ordering-dependent through the decreasing fill terms. 

It is also interesting to remark that the previous expression can be understood in probabilistic terms. Indeed, given that $ \left\langle b_{pa} \right\rangle=p_{ap}$, where $p_{pa}$ are independent link probabilities, the overlap term might be seen as a joint probability of two independent events, divided by a normalizing factor which is the union of independent probabilities. For example, for one pair of animals, the overlap term results in:

\begin{equation}
 \frac{ \sum\limits _{p=1} ^{N_P} \;\left\langle b_{pk} \right\rangle \left\langle b_{pl}\right\rangle }{\sum\limits _{p=1} ^{N_P} \; \left\langle b_{pl} \right\rangle} = \frac{ \sum \limits_{p=1} ^{N_P}\; p_{pk} \; p_{pl}}{\sum \limits_{p=1} ^{N_P} \;p_{pl}} = \frac{ \sum\limits _{p=1} ^{N_P}\; P(p_{pk} \cap \; p_{pl})}{P(p_{1l} \cup p_{2l} \cup .... \; p_{pl} \cup .... \; p_{N_P l})}
\end{equation}

Apart from the first moment, equation \ref{eq5} indicates how to compute the standard deviation. Adapting it to the nestedness measure, we encounter: 

\begin{equation}
\sigma_{\text{NODF}} = \sqrt{\sum_{p=1}^{N_P} \sum_{a=1}^{N_A} \left( \frac{\partial \text{NODF(\textbf{B})}}{\partial b_{pa}} \right)^2 \sigma^2 _{b_{pa}}} \qquad \quad \text{ where }\quad \sigma_{b_{pa}} ^2 = p_{pa}\; (1-p_{pa}) \label{eq7}
\end{equation}

We have introduced that the standard deviation of the probability of links is that of a Bernoulli distribution, since the existence of a link is a binomial process (either there is interaction or there is not). Furthermore, the derivative with respect to a general matrix element $b_{rc}$ (the sub-index \textit{r} stands for rows and \textit{c} stands for columns) can be split into the contributions of plants and of animals:

\begin{equation}
\frac{\partial \; \text{NODF(\textbf{B})}}{\partial b_{rc}} = \frac{\partial \; \text{NODF(\textbf{B})}_{\text{plants}}}{\partial b_{rc}} + \frac{\partial \; \text{NODF(\textbf{B})}_{\text{animals}}}{\partial b_{rc}}
\end{equation}

After deriving, we obtained that: 

\begin{equation}
\frac{\partial \; \text{NODF(\textbf{B})}_{\text{plants}}}{\partial b_{rc}} = \frac{1}{K} \left\lbrace \; \sum_{j = r +1}^{N_P} \! DF_{r j} \; \frac{b_{j c}}{v_j} \; +\; \sum_{i=1}^{r - 1} DF_{ir} \; \frac{b_{i c}}{v_r}\; - \; \sum_{i=1}^{r-1}\sum_{a = 1}^{N_A} DF_{ir} \; \frac{b_{ia} \; b_{ra}}{{v_r}^2} \;\right\rbrace 
\end{equation}

\begin{equation}
\frac{\partial \; \text{NODF(\textbf{B})}_{\text{animals}}}{\partial b_{rc}} =  \frac{1}{K} \left\lbrace \; \sum_{l=c+1}^{N_A} \! DF_{cl} \; \frac{b_{rl}}{h_l} \; + \; \sum_{k=1}^{c-1} DF_{kc} \; \frac{b_{rk}}{h_c} \; - \; \sum_{k = 1}^{c-1} \sum_{p=1}^{N_P} \; DF_{k c} \; \frac{b_{pk} \; b_{p c}}{{h_c}^2} \;\right\rbrace 
\end{equation}

These novel estimations of the first and second moments can then be used to provide a randomized measure of nestedness $\left\langle \text{NODF(\textbf{B})} \right\rangle ^{*}$ together with its statistical significance. 

\subsection*{S3. Nestedness statistical measures using the \textit{largest eigenvalue radius}}

The largest eigenvalue radius was recently proposed by Staniczenko et al.~\cite{staniczenko2013ghost} as an alternative measure for nestedness that directly relies on the spectral properties of the adjacency matrix. The fact that it involves finding the \textit{maximum} eigenvalue entails that we lack an analytical and derivable expression for it. This, in turn, means that we can not derive for this metric the analytical expressions of Eqs.~\ref{eq4}-\ref{eq5} in SI2. Therefore, in order to calculate the statistical properties of our interest, we produced a sample of the statistical ensemble and algorithmically computed the distribution of the largest eigenvalue radius. 

In detail, we produced $10^4$ networks, sampled using the link probabilities in $\left\langle \textbf{B$^{*}$} \right\rangle$ (see Eq.~10, SI1). Then we computed the largest eigenvalue radius, which we call $\rho(\lambda)$, using the $R$ package \textit{rARPACK}~\cite{rarpack}. Finally we calculated the average and the standard deviation of the resulting distribution. As shown in Fig.~\ref{fig3}, the results are fully compatible with those found using the NODF nestedness metric. See also Tables.~\ref{table1} and \ref{table2} for details of percentages.
   
\begin{figure}[!tb]
\centering
\includegraphics[width=0.6\columnwidth]{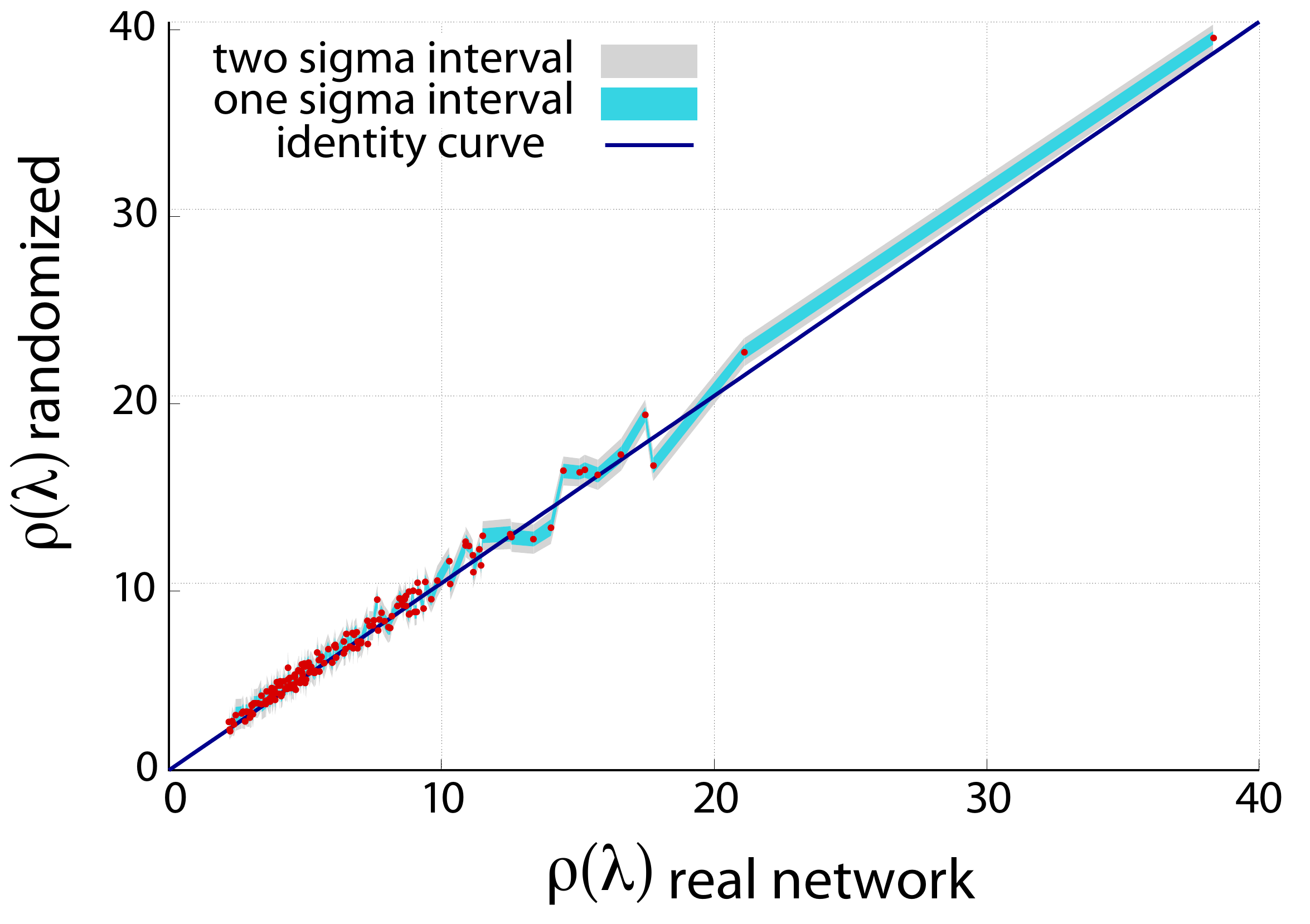} 
\caption{Comparison between the real observation of the largest eigenvalue radius $\rho(\lambda)$ and the average estimation in the statistical ensemble, for the 167 networks of our study.} \label{fig3}
\end{figure}

\begin{center} 
\begin{table}[!tb]
\centering
\begin{tabular}{|c|c|c|c|c|} 
\hline
\multicolumn{2}{|c|}{fraction of ntws with $\vert$z-score$\vert$ $\leq$ 1} & \multicolumn{2}{c|}{fraction of ntws with $\vert$z-score$\vert$ $\leq$ 2} \\ 
\hline 
\hline
$\quad$ 84 out of 167 $\quad$ & 50.3\% & $\quad$ 149 out of 167 $\quad$ & 89.2\%  \\ 
 \hline 
\end{tabular}
\caption{Fraction of networks whose discrepancy between the real and randomized nestedness is less or equal than one or two sigma.}
\label{table1}
\end{table}
\end{center} 

\begin{center} 
\begin{table}[!tb]
\centering
\begin{tabular}{|c|c|c|c|c|} 
\hline
\multicolumn{2}{|c|}{fraction of ntws with corrected $\vert$z-score$\vert$ $\leq$ 1} & \multicolumn{2}{c|}{fraction of ntws with corrected $\vert$z-score$\vert$ $\leq$ 2} \\ 
\hline 
\hline
$\quad$ 151 out of 167 $\quad$ & 90.4\% & $\quad$ 161 out of 167 $\quad$ & 96.4\%  \\ 
 \hline 
\end{tabular}
\caption{After performing the multiple test correction using the false discovery rate method (see Methods summary in Main text), fraction of networks whose discrepancy between the real and randomized nestedness is less or equal than one or two sigma.}
\label{table2}
\end{table}
\end{center}

\begin{figure}[!tb]
\centering
\includegraphics[width=0.6\columnwidth]{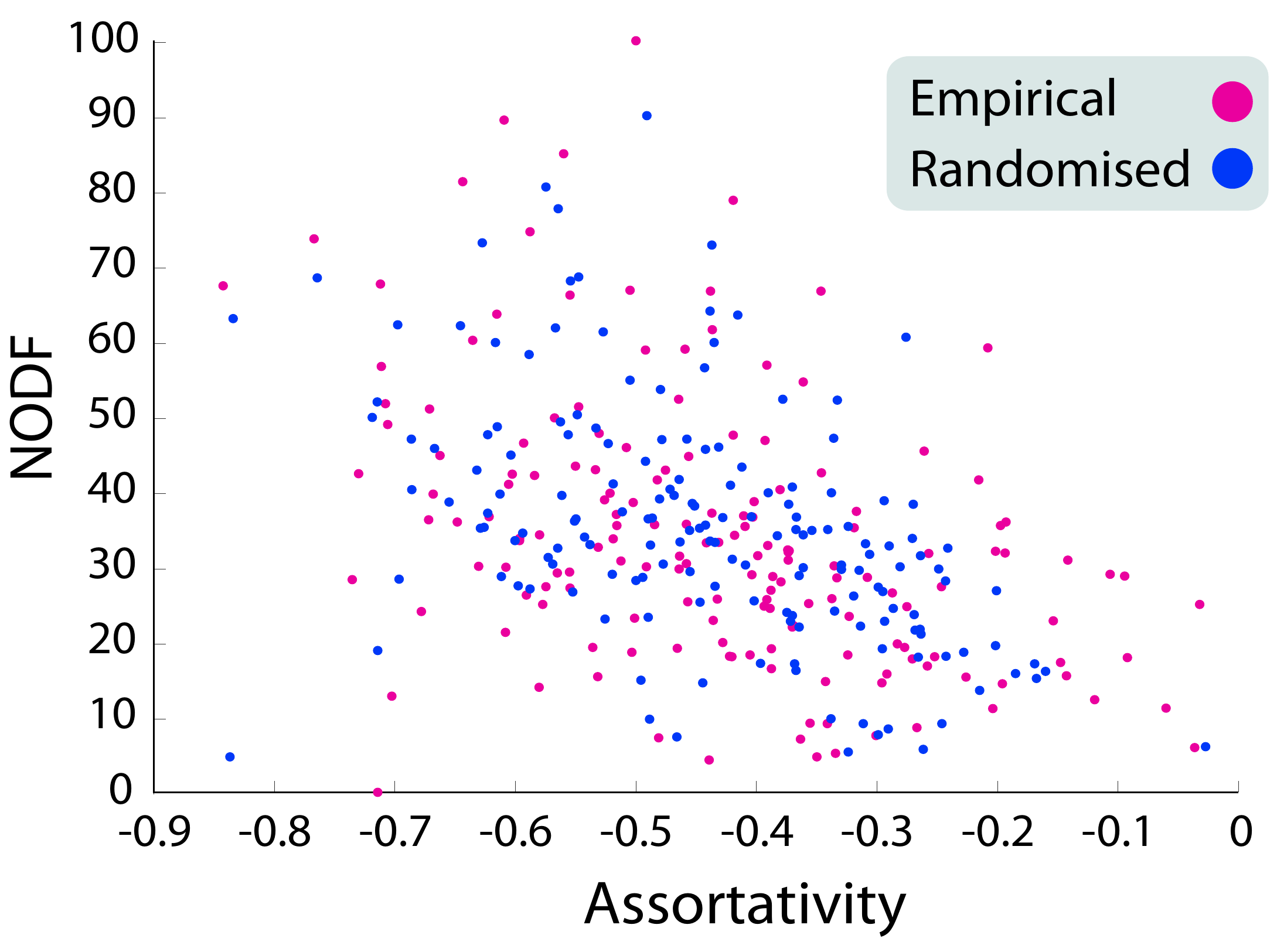} 
\caption{Relation between the degree assortativity and the nestedness for the real network (in red) and the average measure for the randomised case (in blue).} \label{fig5}
\end{figure}

\subsection*{S4. Assortativity statistical measures}

Assortativity is a network feature that quantifies to what extent nodes tend to match other nodes that are similar (or dissimilar) to them. Here, we particularly used the notion of degree assortativity, which means that \textit{similarity} is labelled by the degree. We followed the definition proposed by Newman~\cite{newman2002assortative}, which consists of a normalized correlation coefficient between degrees. This eventually corresponds to the \textit{Pearson correlation coefficient} denoted by \textit{r}, such that $r=-1$ indicates perfect disassortativity, $r=0$ no correlation at all and $r=1$ maximum assortativity.  

In order to compute the statistical properties of this quantity, we produced for each ensemble a sampling made up by $10^4$ networks. We then measured computationally the assortativity of each sampled network using the \textit{assortativity\_degree} function from the \textit{igraph} package in $R$~\cite{assortat}. Finally this allowed us to calculate the first and second moments of the assortativity for each ensemble in our set.

\subsection*{S5. Data}

In our study we analyzed 167 real interaction networks from the \textit{Web of Life} dataset~\cite{weboflife}. This set consists of 133 plant-pollinator communities [references from 47 to 83], 30 seed-dispersal [references from 84 to 104] and 4 plant-ant [references from 105 to 108]. 

Data sometimes included information about link's weight, but we converted all networks to binary matrices.  

\nocite{arroyo1982community} \nocite{barrett1987reproductive} \nocite{clements1923experimental} \nocite{dicks2002compartmentalization} \nocite{dupont2003structure} \nocite{elberling1999structure} \nocite{olesen2002invasion} \nocite{ollerton2003pollination} \nocite{hocking1968insect} \nocite{petanidou1991pollination} \nocite{herrera1988pollination} \nocite{memmott1999structure} \nocite{inouye1988pollination} \nocite{kevan2002high} \nocite{kato1990insect} \nocite{medan2002plant} \nocite{mosquin1967observations} \nocite{motten1984pollination} \nocite{mcmullen1993flower} \nocite{primack1983insect} \nocite{ramirez1992pollination} \nocite{ramirez1989biologia} \nocite{schemske1978flowering} \nocite{small1976insect} \nocite{smith2005diversity} \nocite{percival1974floral} \nocite{montero2005ecology} \nocite{ingvergsen2006plant} \nocite{philipp2006structure} \nocite{montero2005ecology} \nocite{kato2000anthophilous} \nocite{lundgren2005dense} \nocite{bundgaard2003tidslig} \nocite{dupont2009ecological} \nocite{bek2006pollination} \nocite{stald2003struktur} \nocite{vazquez2002interactions} \nocite{witt1998} \nocite{yamazaki2003flowering} 
  
\nocite{baird1980selection}\nocite{beehler1983frugivory}\nocite{carlo2003avian}\nocite{crome1975ecology}\nocite{frost1980fruit}\nocite{galetti2013fruit}\nocite{snow1971feeding}\nocite{snow1988feeding}\nocite{hamann1999interactions}\nocite{jordano1985ciclo}\nocite{kantak1979observations}\nocite{lambert1989fig}\nocite{tutin1997primate}\nocite{mack1996notes}\nocite{wheelwright1984tropical}\nocite{silva2002patterns}\nocite{noma1997annual}\nocite{guitian1983relaciones}\nocite{heleno2013integration}\nocite{poulin1999interspecific}\nocite{schleuning2011specialization}

\nocite{bluthgen2004bottom}\nocite{davidson1989competition}\nocite{davidson1991symbiosis}\nocite{fonseca1996asymmetries}  

\bibliographystyle{ieeetr} 

\bibliography{ref}

\begin{thebibliography}{100}

\bibitem{bascompte2003nested}
J.~Bascompte, P.~Jordano, C.~J. Meli{\'a}n, and J.~M. Olesen, ``The nested
  assembly of plant--animal mutualistic networks,'' {\em Proceedings of the
  National Academy of Sciences}, vol.~100, no.~16, pp.~9383--9387, 2003.

\bibitem{bastolla2009architecture}
U.~Bastolla, M.~A. Fortuna, A.~Pascual-Garc{\'\i}a, A.~Ferrera, B.~Luque, and
  J.~Bascompte, ``The architecture of mutualistic networks minimizes
  competition and increases biodiversity,'' {\em Nature}, vol.~458, no.~7241,
  pp.~1018--1020, 2009.

\bibitem{thebault2010stability}
E.~Th{\'e}bault and C.~Fontaine, ``Stability of ecological communities and the
  architecture of mutualistic and trophic networks,'' {\em Science}, vol.~329,
  no.~5993, pp.~853--856, 2010.

\bibitem{rohr2014structural}
R.~P. Rohr, S.~Saavedra, and J.~Bascompte, ``On the structural stability of
  mutualistic systems,'' {\em Science}, vol.~345, no.~6195, p.~1253497, 2014.

\bibitem{burgos2007nestedness}
E.~Burgos, H.~Ceva, R.~P. Perazzo, M.~Devoto, D.~Medan, M.~Zimmermann, and
  A.~M. Delbue, ``Why nestedness in mutualistic networks?,'' {\em Journal of
  theoretical biology}, vol.~249, no.~2, pp.~307--313, 2007.

\bibitem{suweis2013emergence}
S.~Suweis, F.~Simini, J.~R. Banavar, and A.~Maritan, ``Emergence of structural
  and dynamical properties of ecological mutualistic networks,'' {\em Nature},
  vol.~500, no.~7463, pp.~449--452, 2013.

\bibitem{valverde2016nestedness}
S.~Valverde, J.~Montoya, L.~Joppa, and R.~Sole, ``Is nestedness in mutualistic
  networks an evolutionary spandrel?,'' {\em ArXiv preprint}.

\bibitem{ulrich2009consumer}
W.~Ulrich, M.~Almeida-Neto, and N.~J. Gotelli, ``A consumer's guide to
  nestedness analysis,'' {\em Oikos}, vol.~118, no.~1, pp.~3--17, 2009.

\bibitem{vazquez2006community}
D.~P. V{\'a}zquez and M.~A. Aizen, ``Community-wide patterns of specialization
  in plant--pollinator interactions revealed by null models,'' {\em
  Plant--pollinator interactions: from specialization to generalization},
  pp.~200--219, 2006.

\bibitem{rezende2007effects}
E.~L. Rezende, P.~Jordano, and J.~Bascompte, ``Effects of phenotypic
  complementarity and phylogeny on the nested structure of mutualistic
  networks,'' {\em Oikos}, vol.~116, no.~11, pp.~1919--1929, 2007.

\bibitem{perazzo2014study}
R.~P. Perazzo, L.~Hern{\'a}ndez, H.~Ceva, E.~Burgos, and J.~I. Alvarez-Hamelin,
  ``Study of the influence of the phylogenetic distance on the interaction
  network of mutualistic ecosystems,'' {\em Physica A: Statistical Mechanics
  and its Applications}, vol.~394, pp.~124--135, 2014.

\bibitem{staniczenko2013ghost}
P.~P. Staniczenko, J.~C. Kopp, and S.~Allesina, ``The ghost of nestedness in
  ecological networks,'' {\em Nature communications}, vol.~4, p.~1391, 2013.

\bibitem{james2012disentangling}
A.~James, J.~W. Pitchford, and M.~J. Plank, ``Disentangling nestedness from
  models of ecological complexity,'' {\em Nature}, vol.~487, no.~7406,
  pp.~227--230, 2012.

\bibitem{jonhson2013factors}
S.~Jonhson, V.~Dom{\'\i}nguez-Garc{\'\i}a, and M.~A. Mu{\~n}oz, ``Factors
  determining nestedness in complex networks,'' {\em PloS one}, vol.~8, no.~9,
  p.~e74025, 2013.

\bibitem{feng2014heterogeneity}
W.~Feng and K.~Takemoto, ``Heterogeneity in ecological mutualistic networks
  dominantly determines community stability,'' {\em Scientific reports},
  vol.~4, p.~5912, 2014.

\bibitem{medan2007analysis}
D.~Medan, R.~P. Perazzo, M.~Devoto, E.~Burgos, M.~G. Zimmermann, H.~Ceva, and
  A.~M. Delbue, ``Analysis and assembling of network structure in mutualistic
  systems,'' {\em Journal of theoretical biology}, vol.~246, no.~3,
  pp.~510--521, 2007.

\bibitem{atmar1993measure}
W.~Atmar and B.~D. Patterson, ``The measure of order and disorder in the
  distribution of species in fragmented habitat,'' {\em Oecologia}, vol.~96,
  no.~3, pp.~373--382, 1993.

\bibitem{olesen2010missing}
J.~M. Olesen, J.~Bascompte, Y.~L. Dupont, H.~Elberling, C.~Rasmussen, and
  P.~Jordano, ``Missing and forbidden links in mutualistic networks,'' {\em
  Proceedings of the Royal Society of London B: Biological Sciences},
  pp.~13--71, 2010.

\bibitem{joppa2010nestedness}
L.~N. Joppa, J.~M. Montoya, R.~Sol{\'e}, J.~Sanderson, and S.~L. Pimm, ``On
  nestedness in ecological networks,'' {\em Evolutionary Ecology Research},
  vol.~12, no.~1, pp.~35--46, 2010.

\bibitem{squartini2011analytical}
T.~Squartini and D.~Garlaschelli, ``Analytical maximum-likelihood method to
  detect patterns in real networks,'' {\em New Journal of Physics}, vol.~13,
  no.~8, p.~083001, 2011.

\bibitem{saracco2015randomizing}
F.~Saracco, R.~Di~Clemente, A.~Gabrielli, and T.~Squartini, ``Randomizing
  bipartite networks: the case of the world trade web,'' {\em Scientific
  Reports}, vol.~5, no.~10595, 2015.

\bibitem{garlaschelli2008maximum}
D.~Garlaschelli and M.~I. Loffredo, ``Maximum likelihood: extracting unbiased
  information from complex networks,'' {\em Physical Review E}, vol.~78, no.~1,
  p.~015101, 2008.

\bibitem{inoue1990insect}
T.~Inoue, M.~Kato, T.~Kakutani, T.~Suka, and T.~Itino, ``Insect-flower
  relationship in the temperate deciduous forest of kibune, kyoto: an overview
  of the flowering phenology and the seasonal pattern of insect visits,'' 1990.

\bibitem{ulrich2007null}
W.~Ulrich and N.~J. Gotelli, ``Null model analysis of species nestedness
  patterns,'' {\em Ecology}, vol.~88, no.~7, pp.~1824--1831, 2007.

\bibitem{jordano2003invariant}
P.~Jordano, J.~Bascompte, and J.~M. Olesen, ``Invariant properties in
  coevolutionary networks of plant--animal interactions,'' {\em Ecology
  Letters}, vol.~6, no.~1, pp.~69--81, 2003.

\bibitem{memmott2004tolerance}
J.~Memmott, N.~M. Waser, and M.~V. Price, ``Tolerance of pollination networks
  to species extinctions,'' {\em Proceedings of the Royal Society of London B:
  Biological Sciences}, vol.~271, no.~1557, pp.~2605--2611, 2004.

\bibitem{gotelli2001swap}
N.~J. Gotelli and G.~L. Entsminger, ``Swap and fill algorithms in null model
  analysis: rethinking the knight's tour,'' {\em Oecologia}, vol.~129, no.~2,
  pp.~281--291, 2001.

\bibitem{johnson2010entropic}
S.~Johnson, J.~J. Torres, J.~Marro, and M.~A. Muñoz, ``Entropic origin of
  disassortativity in complex networks,'' {\em Physical review letters},
  vol.~104, no.~10, p.~108702, 2010.

\bibitem{takemoto2010nested}
K.~Takemoto and M.~Arita, ``Nested structure acquired through simple
  evolutionary process,'' {\em Journal of theoretical biology}, vol.~264,
  no.~3, pp.~782--786, 2010.

\bibitem{johnson2000generalization}
S.~D. Johnson and K.~E. Steiner, ``Generalization versus specialization in
  plant pollination systems,'' {\em Trends in Ecology \& Evolution}, vol.~15,
  no.~4, pp.~140--143, 2000.

\bibitem{bronstein2006evolution}
J.~L. Bronstein, R.~Alarc{\'o}n, and M.~Geber, ``The evolution of plant--insect
  mutualisms,'' {\em New Phytologist}, vol.~172, no.~3, pp.~412--428, 2006.

\bibitem{bronstein2001costs}
J.~L. Bronstein, ``The costs of mutualism,'' {\em American Zoologist}, vol.~41,
  no.~4, pp.~825--839, 2001.

\bibitem{benjamini1995controlling}
Y.~Benjamini and Y.~Hochberg, ``Controlling the false discovery rate: a
  practical and powerful approach to multiple testing,'' {\em Journal of the
  royal statistical society. Series B (Methodological)}, pp.~289--300, 1995.

\bibitem{park2004statistical}
J.~Park and M.~E. Newman, ``Statistical mechanics of networks,'' {\em Physical
  Review E}, vol.~70, no.~6, p.~066117, 2004.

\bibitem{corana1987minimizing}
A.~Corana, M.~Marchesi, C.~Martini, and S.~Ridella, ``Minimizing multimodal
  functions of continuous variables with the “simulated annealing”
  algorithm corrigenda for this article is available here,'' {\em ACM
  Transactions on Mathematical Software (TOMS)}, vol.~13, no.~3, pp.~262--280,
  1987.

\bibitem{goffe1994global}
W.~L. Goffe, G.~D. Ferrier, and J.~Rogers, ``Global optimization of statistical
  functions with simulated annealing,'' {\em Journal of econometrics}, vol.~60,
  no.~1-2, pp.~65--99, 1994.

\bibitem{goffe1996simann}
W.~L. Goffe {\em et~al.}, ``Simann: A global optimization algorithm using
  simulated annealing,'' {\em Studies in Nonlinear Dynamics and Econometrics},
  vol.~1, no.~3, pp.~169--176, 1996.

\bibitem{metropolis1953equation}
N.~Metropolis, A.~W. Rosenbluth, M.~N. Rosenbluth, A.~H. Teller, and E.~Teller,
  ``Equation of state calculations by fast computing machines,'' {\em The
  journal of chemical physics}, vol.~21, no.~6, pp.~1087--1092, 1953.

\bibitem{more1980user}
J.~J. Mor{\'e}, B.~S. Garbow, and K.~E. Hillstrom, ``User guide for
  minpack-1,'' tech. rep., CM-P00068642, 1980.

\bibitem{minpack}
J.~J.~M. Burton S.~Garbow, Kenneth E.~Hillstrom, {\em MINPACK libary source
  code, subroutine HYBRD}.
\newblock Available at:
  \url{https://www.math.utah.edu/software/minpack/minpack/hybrd.html}.

\bibitem{almeida2008consistent}
M.~Almeida-Neto, P.~Guimar{\~a}es, P.~R. Guimar{\~a}es, R.~D. Loyola, and
  W.~Ulrich, ``A consistent metric for nestedness analysis in ecological
  systems: reconciling concept and measurement,'' {\em Oikos}, vol.~117, no.~8,
  pp.~1227--1239, 2008.

\bibitem{petanidou1991pollination}
T.~Petanidou, ``Pollination ecology in a phryganic ecosystem,'' 1991.

\bibitem{rarpack}
{\em rARPACK: Solvers for Large Scale Eigenvalue and SVD Problems}.
\newblock Available at: \url{https://CRAN.R-project.org/package=rARPACK }.

\bibitem{newman2002assortative}
M.~E. Newman, ``Assortative mixing in networks,'' {\em Physical review
  letters}, vol.~89, no.~20, p.~208701, 2002.

\bibitem{assortat}
G.~Csardi, {\em Assortativity in R igraph package}.
\newblock Available at: \url{http://igraph.org/r/doc/assortativity.html}.

\bibitem{weboflife}
Bascompte Lab, {\em Web of Life, ecological networks database}.
\newblock Available at \url{http://www.web-of-life.es/}.

\bibitem{arroyo1982community}
M.~T.~K. Arroyo, R.~Primack, and J.~Armesto, ``Community studies in pollination
  ecology in the high temperate andes of central chile. i. pollination
  mechanisms and altitudinal variation,'' {\em American journal of botany},
  pp.~82--97, 1982.

\bibitem{barrett1987reproductive}
S.~C. Barrett and K.~Helenurm, ``The reproductive biology of boreal forest
  herbs. i. breeding systems and pollination,'' {\em Canadian Journal of
  Botany}, vol.~65, no.~10, pp.~2036--2046, 1987.

\bibitem{clements1923experimental}
F.~E. Clements and F.~L. Long, {\em Experimental pollination: an outline of the
  ecology of flowers and insects}.
\newblock No.~336, Carnegie institution of Washington, 1923.

\bibitem{dicks2002compartmentalization}
L.~Dicks, S.~Corbet, and R.~Pywell, ``Compartmentalization in plant--insect
  flower visitor webs,'' {\em Journal of Animal Ecology}, vol.~71, no.~1,
  pp.~32--43, 2002.

\bibitem{dupont2003structure}
Y.~L. Dupont, D.~M. Hansen, and J.~M. Olesen, ``Structure of a
  plant--flower-visitor network in the high-altitude sub-alpine desert of
  tenerife, canary islands,'' {\em Ecography}, vol.~26, no.~3, pp.~301--310,
  2003.

\bibitem{elberling1999structure}
H.~Elberling and J.~M. Olesen, ``The structure of a high latitude plant-flower
  visitor system: the dominance of flies,'' {\em Ecography}, vol.~22, no.~3,
  pp.~314--323, 1999.

\bibitem{olesen2002invasion}
J.~M. Olesen, L.~I. Eskildsen, and S.~Venkatasamy, ``Invasion of pollination
  networks on oceanic islands: importance of invader complexes and endemic
  super generalists,'' {\em Diversity and Distributions}, vol.~8, no.~3,
  pp.~181--192, 2002.

\bibitem{ollerton2003pollination}
J.~Ollerton, S.~D. Johnson, L.~Cranmer, and S.~Kellie, ``The pollination
  ecology of an assemblage of grassland asclepiads in south africa,'' {\em
  Annals of Botany}, vol.~92, no.~6, pp.~807--834, 2003.

\bibitem{hocking1968insect}
B.~Hocking, ``Insect-flower associations in the high arctic with special
  reference to nectar,'' {\em Oikos}, pp.~359--387, 1968.

\bibitem{herrera1988pollination}
J.~Herrera, ``Pollination relationships in southern spanish mediterranean
  shrublands,'' {\em The Journal of Ecology}, pp.~274--287, 1988.

\bibitem{memmott1999structure}
J.~Memmott, ``The structure of a plant-pollinator food web,'' {\em Ecology
  letters}, vol.~2, no.~5, pp.~276--280, 1999.

\bibitem{inouye1988pollination}
D.~W. Inouye and G.~H. Pyke, ``Pollination biology in the snowy mountains of
  australia: comparisons with montane colorado, usa,'' {\em Austral Ecology},
  vol.~13, no.~2, pp.~191--205, 1988.

\bibitem{kevan2002high}
P.~G. Kevan, ``High arctic insect-flower relations the interrelationships of
  arthropods and flowers at lake hazen, ellesmere island, nwt, canada.,'' 2002.

\bibitem{kato1990insect}
M.~Kato, T.~Kakutani, T.~Inoue, and T.~Itino, ``Insect-flower relationship in
  the primary beech forest of ashu, kyoto: an overview of the flowering
  phenology and the seasonal pattern of insect visits,'' 1990.

\bibitem{medan2002plant}
D.~Medan, N.~H. Montaldo, M.~Devoto, A.~Mantese, V.~Vasellati, G.~G. Roitman,
  and N.~H. Bartoloni, ``Plant-pollinator relationships at two altitudes in the
  andes of mendoza, argentina,'' {\em Arctic, Antarctic, and Alpine Research},
  pp.~233--241, 2002.

\bibitem{mosquin1967observations}
T.~Mosquin and J.~Martin, ``Observations on the pollination biology of plants
  on melville island, nwt, canada,'' {\em Canadian Field Naturalist}, vol.~81,
  pp.~201--205, 1967.

\bibitem{motten1984pollination}
``Pollination ecology of the spring wildflower community in the deciduous
  forests of piedmont north carolina,''

\bibitem{mcmullen1993flower}
C.~McMullen, ``Flower-visiting insects of the galapagos islands,'' {\em
  Pan-Pacific Entomologist}, vol.~69, no.~1, pp.~95--106, 1993.

\bibitem{primack1983insect}
R.~B. Primack, ``Insect pollination in the new zealand mountain flora,'' {\em
  New Zealand Journal of Botany}, vol.~21, no.~3, pp.~317--333, 1983.

\bibitem{ramirez1992pollination}
N.~Ramirez and Y.~Brito, ``Pollination biology in a palm swamp community in the
  venezuelan central plains,'' {\em Botanical Journal of the Linnean Society},
  vol.~110, no.~4, pp.~277--302, 1992.

\bibitem{ramirez1989biologia}
N.~Ramirez, ``Biologia de polinizacion en una comunidad arbustiva tropical de
  la alta guayana venezolana,'' {\em Biotropica}, pp.~319--330, 1989.

\bibitem{schemske1978flowering}
D.~W. Schemske, M.~F. Willson, M.~N. Melampy, L.~J. Miller, L.~Verner, K.~M.
  Schemske, and L.~B. Best, ``Flowering ecology of some spring woodland
  herbs,'' {\em Ecology}, vol.~59, no.~2, pp.~351--366, 1978.

\bibitem{small1976insect}
E.~Small, ``Insect pollinators of the mer bleue peat bog of ottawa,'' {\em
  Canadian field-naturalist}, 1976.

\bibitem{smith2005diversity}
C.~Smith-Ramirez, P.~Martinez, M.~Nunez, C.~Gonz{\'a}lez, and J.~J. Armesto,
  ``Diversity, flower visitation frequency and generalism of pollinators in
  temperate rain forests of chilo{\'e} island, chile,'' {\em Botanical Journal
  of the Linnean Society}, vol.~147, no.~4, pp.~399--416, 2005.

\bibitem{percival1974floral}
M.~Percival, ``Floral ecology of coastal scrub in southeast jamaica,'' {\em
  Biotropica}, pp.~104--129, 1974.

\bibitem{montero2005ecology}
A.~Montero, {\em The Ecology of Three Pollination Networks}.
\newblock PhD thesis, MSc thesis, Aarhus University, Aarhus, 2005.

\bibitem{ingvergsen2006plant}
T.~Ingversen, {\em Plant–pollinator interactions on Jamaica and Dominica: The
  centrality, asymmetry and modularity of networks}.
\newblock PhD thesis, MSc thesis, Aarhus University, Aarhus, 2006.

\bibitem{philipp2006structure}
M.~Philipp, J.~B{\"o}cher, H.~R~Siegismund, and L.~R~Nielsen, ``Structure of a
  plant-pollinator network on a pahoehoe lava desert of the gal{\'a}pagos
  islands,'' {\em Ecography}, vol.~29, no.~4, pp.~531--540, 2006.

\bibitem{kato2000anthophilous}
M.~Kato, ``Anthophilous insect community and plant-pollinator interactions on
  amami islands in the ryukyu archipelago, japan (original paper),'' 2000.

\bibitem{lundgren2005dense}
R.~Lundgren and J.~M. Olesen, ``The dense and highly connected world of
  greenland's plants and their pollinators,'' {\em Arctic, Antarctic, and
  Alpine Research}, vol.~37, no.~4, pp.~514--520, 2005.

\bibitem{bundgaard2003tidslig}
M.~Bundgaard, {\em Tidslig og rumlig variation i et
  plante-best{\o}vernetv{\ae}rk}.
\newblock PhD thesis, MSc. thesis, Aarhus University, Aarhus., 2003.

\bibitem{dupont2009ecological}
Y.~L. Dupont and J.~M. Olesen, ``Ecological modules and roles of species in
  heathland plant--insect flower visitor networks,'' {\em Journal of Animal
  Ecology}, vol.~78, no.~2, pp.~346--353, 2009.

\bibitem{bek2006pollination}
S.~Bek, {\em A pollination network from a Danish forest meadow}.
\newblock PhD thesis, MSc. thesis, Aarhus University, Aarhus., 2006.

\bibitem{stald2003struktur}
L.~Stald, A.~Valido, and J.~Olesen, {\em Struktur og dynamik i rum og tid at et
  best{\o}vningsnetv{\ae}rk p{\aa} Tenerife, De Kanariske {\O}er}.
\newblock PhD thesis, MSc-thesis, University of Aarhus, Denmark, 2003.

\bibitem{vazquez2002interactions}
D.~P. V{\'a}zquez, ``Interactions among introduced ungulates, plants, and
  pollinators: a field study in the temperate forest of the southern andes,''
  2002.

\bibitem{witt1998}
W.~P.
\newblock PhD thesis, BSc thesis, University of Aarhus, Denmark, 1998.

\bibitem{yamazaki2003flowering}
K.~Yamazaki and M.~Kato, ``Flowering phenology and anthophilous insect
  community in a grassland ecosystem at mt. yufu, western japan,'' 2003.

\bibitem{baird1980selection}
J.~W. Baird, ``The selection and use of fruit by birds in an eastern forest,''
  {\em The Wilson Bulletin}, pp.~63--73, 1980.

\bibitem{beehler1983frugivory}
B.~Beehler, ``Frugivory and polygamy in birds of paradise,'' {\em The Auk},
  pp.~1--12, 1983.

\bibitem{carlo2003avian}
T.~A. Carlo, J.~A. Collazo, and M.~J. Groom, ``Avian fruit preferences across a
  puerto rican forested landscape: pattern consistency and implications for
  seed removal,'' {\em Oecologia}, vol.~134, no.~1, pp.~119--131, 2003.

\bibitem{crome1975ecology}
F.~Crome, ``The ecology of fruit pigeons in tropical northern queensland.,''
  {\em Wildlife Research}, vol.~2, no.~2, pp.~155--185, 1975.

\bibitem{frost1980fruit}
P.~Frost, ``Fruit-frugivore interactions in a south african coastal dune
  forest,'' 1980.

\bibitem{galetti2013fruit}
M.~Galetti and M.~A. Pizo, ``Fruit eating by birds in a forest fragment in
  southeastern brazil.,'' {\em Revista Brasileira de Ornitologia-Brazilian
  Journal of Ornithology}, vol.~4, no.~5, p.~9, 2013.

\bibitem{snow1971feeding}
B.~K. Snow and D.~Snow, ``The feeding ecology of tanagers and honeycreepers in
  trinidad,'' {\em The Auk}, vol.~88, no.~2, pp.~291--322, 1971.

\bibitem{snow1988feeding}
B.~K. Snow and D.~Snow, ``Birds and berries,'' 1988.

\bibitem{hamann1999interactions}
A.~Hamann and E.~Curio, ``Interactions among frugivores and fleshy fruit trees
  in a philippine submontane rainforest,'' {\em Conservation Biology}, vol.~13,
  no.~4, pp.~766--773, 1999.

\bibitem{jordano1985ciclo}
P.~Jordano, ``El ciclo anual de los paseriformes frug{\'\i}voros en el matorral
  mediterr{\'a}neo del sur de espa{\~n}a: importancia de su invernada y
  variaciones interanuales,'' {\em Ardeola}, vol.~32, no.~1, pp.~69--94, 1985.

\bibitem{kantak1979observations}
G.~E. Kantak, ``Observations on some fruit-eating birds in mexico,'' {\em The
  Auk}, vol.~96, no.~1, pp.~183--186, 1979.

\bibitem{lambert1989fig}
F.~Lambert, ``Fig-eating by birds in a malaysian lowland rain forest,'' {\em
  Journal of Tropical Ecology}, vol.~5, no.~4, pp.~401--412, 1989.

\bibitem{tutin1997primate}
C.~E. Tutin, R.~M. Ham, L.~J. White, and M.~J. Harrison, ``The primate
  community of the lop{\'e} reserve, gabon: diets, responses to fruit scarcity,
  and effects on biomass,'' {\em American Journal of Primatology}, vol.~42,
  no.~1, pp.~1--24, 1997.

\bibitem{mack1996notes}
A.~L. Mack and D.~D. Wright, ``Notes on occurrence and feeding of birds at
  crater mountain biological research station, papua new guinea,'' {\em Emu},
  vol.~96, no.~2, pp.~89--101, 1996.

\bibitem{wheelwright1984tropical}
N.~T. Wheelwright, W.~A. Haber, K.~G. Murray, and C.~Guindon, ``Tropical
  fruit-eating birds and their food plants: a survey of a costa rican lower
  montane forest,'' {\em Biotropica}, pp.~173--192, 1984.

\bibitem{silva2002patterns}
W.~Silva, ``Patterns of fruit-frugivore interactions in two atlantic forest
  bird communities of south-eastern brazil: implications for conservation,''
  {\em Seed dispersal and frugivory: Ecology, evolution and conservation},
  pp.~423--435, 2002.

\bibitem{noma1997annual}
N.~Noma, ``Annual fluctuations of sapfruits production and synchronization
  within and inter species in a warm temperate forest on yakushima island,''
  {\em Tropics}, vol.~6, pp.~441--449, 1997.

\bibitem{guitian1983relaciones}
J.~Guiti{\'a}n, ``Relaciones entre los frutos y los passeriformes en un bosque
  montano de la cordillera cant{\'a}brica occidental,'' {\em Univ. Santiago,
  Spain}, 1983.

\bibitem{heleno2013integration}
R.~H. Heleno, J.~A. Ramos, and J.~Memmott, ``Integration of exotic seeds into
  an azorean seed dispersal network,'' {\em Biological Invasions}, vol.~15,
  no.~5, pp.~1143--1154, 2013.

\bibitem{poulin1999interspecific}
B.~Poulin, S.~J. Wright, G.~Lefebvre, and O.~Calder{\'o}n, ``Interspecific
  synchrony and asynchrony in the fruiting phenologies of congeneric
  bird-dispersed plants in panama,'' {\em Journal of Tropical Ecology},
  vol.~15, no.~2, pp.~213--227, 1999.

\bibitem{schleuning2011specialization}
M.~Schleuning, N.~Bl{\"u}thgen, M.~Fl{\"o}rchinger, J.~Braun, H.~M. Schaefer,
  and K.~B{\"o}hning-Gaese, ``Specialization and interaction strength in a
  tropical plant--frugivore network differ among forest strata,'' {\em
  Ecology}, vol.~92, no.~1, pp.~26--36, 2011.

\bibitem{bluthgen2004bottom}
N.~Bl{\"u}thgen, N.~E~Stork, and K.~Fiedler, ``Bottom-up control and
  co-occurrence in complex communities: Honeydew and nectar determine a
  rainforest ant mosaic,'' {\em Oikos}, vol.~106, no.~2, pp.~344--358, 2004.

\bibitem{davidson1989competition}
D.~W. Davidson, R.~R. Snelling, and J.~T. Longino, ``Competition among ants for
  myrmecophytes and the significance of plant trichomes,'' {\em Biotropica},
  pp.~64--73, 1989.

\bibitem{davidson1991symbiosis}
D.~Davidson and B.~Fisher, ``Symbiosis of ants with cecropia as a function of
  light regime,'' {\em Huxley, C, R,, Cutler, D, F ed (s). Ant-plant
  interactions. Oxford Univ. Press: Oxford, etc}, pp.~289--309, 1991.

\bibitem{fonseca1996asymmetries}
C.~R. Fonseca and G.~Ganade, ``Asymmetries, compartments and null interactions
  in an amazonian ant-plant community,'' {\em Journal of Animal Ecology},
  pp.~339--347, 1996.

\end{thebibliography}

\end{document}